\documentclass[aps,amsfonts,amsmath,prd,preprint,nofootinbib]{revtex4}
\newcommand{\beq}{\begin{equation}}
\newcommand{\eeq}{\end{equation}}
\usepackage{graphicx}

\begin{document}

\title{Flyover vacuum decay}

\author{Jose J. Blanco-Pillado{$^{1,2}$}, Heling Deng{$^3$} and Alexander Vilenkin{$^3$}}

\affiliation{ $^1$ Department of Theoretical Physics, UPV/EHU, 48080, Bilbao, Spain \\
$^2$ IKERBASQUE, Basque Foundation for Science, 48011, Bilbao, Spain \\
$^3$ Institute of Cosmology, Department of Physics and Astronomy,\\ 
Tufts University, Medford, MA 02155, USA}

\begin{abstract}

We use analytic estimates and numerical simulations to explore the stochastic 
approach to vacuum decay.  According to this approach, the time derivative of a 
scalar field, which is in a local vacuum state, develops a large fluctuation and 
the field ``flies over'' a potential barrier to another vacuum.  The probability 
distribution for the initial fluctuation is found quantum mechanically, while the 
subsequent nonlinear evolution is determined by classical dynamics. We find in 
a variety of cases that the rate of such flyover transitions has the same parametric 
form as that of tunneling transitions calculated using the instanton method, differing 
only by a numerical factor O(1) in the exponent.  An important exception is an 
"upward"  transition from a de Sitter vacuum to a higher-energy de Sitter vacuum 
state.  The rate of flyover transitions in this case is parametrically different and 
can be many orders of magnitude higher than tunneling.  This result is in conflict 
with the conventional picture of quantum de Sitter space as a thermal state.  Our 
numerical simulations indicate that the dynamics of bubble nucleation in flyover 
transitions is rather different from the standard picture.  The difference is 
especially strong for thin-wall bubbles in flat space, where the transition region 
oscillates between true and false vacuum until a true vacuum shell is formed 
which expands both inwards and outwards, and for upward de Sitter transitions, 
where the inflating new vacuum region is contained inside of a black hole.

\end{abstract}

\maketitle

\section{Introduction}

The stochastic approach to vacuum decay is based on a heuristic picture, first suggested 
by Linde et al \cite{Ellis:1990bv,Linde:1991sk} and further 
developed in Refs.~\cite{Brown:2011ry,Braden:2018tky,Huang, Hertzberg:2019wgx}.
According to this picture, the field $\phi$, which is initially in the false vacuum state, 
develops a large quantum fluctuation which then takes it over the barrier to the true 
vacuum.  The probability distribution for the initial fluctuations is found quantum 
mechanically, with the field $\phi$ treated as a free quantum field in the false vacuum, 
but the subsequent nonlinear evolution of the fluctuation is determined using the 
classical field equations.  This approach was applied to several examples and 
reproduced by order of magnitude the exponent in the tunneling rate obtained 
by the standard Coleman's instanton method \cite{Coleman:1977py}.  Braden et 
al \cite{Braden:2018tky} used the stochastic approach to study vacuum decay 
numerically in $(1+1)$ dimensions and found the decay rate in a quantitative 
agreement with the standard calculation.  This raises the question of whether 
this approach describes a different channel of vacuum decay or just gives an 
approximate method of calculating the decay rate.

This question is particularly interesting in the multiverse context, where quantum 
transitions between different vacua can be significantly affected by gravity.  An 
elegant instanton formalism to describe vacuum decay in the presence of gravity 
was developed by Coleman and De Luccia (CdL) \cite{Coleman:1980aw}.  According 
to this formalism, the decay rate of a metastable de Sitter (dS) vacuum is given 
by\footnote{Here and below we disregard prefactors in the expressions for the 
tunneling rate.}
\beq
\kappa_{CdL} \sim e^{-I-S_p}.   
\label{CdL}
\eeq
Here, $I<0$ is the instanton action, $S_p=\pi/H_p^2$ is the dS entropy and 
$H_p$ is the expansion rate of the parent (decaying) vacuum\footnote{Throughout 
this paper we will take $M_p=1$.}.  Lee and Weinberg \cite{Lee:1987qc} have 
pointed out that this formula should apply not only to downward but also to 
upward transitions, where the daughter vacuum has a higher energy density 
than the parent vacuum. Analytic continuation of the instanton to the Lorentzian 
regime indicates that the initial bubble size in this case is greater than the parent 
vacuum horizon $H_p^{-1}$.

CdL instantons may not exist when the potential barrier between the two vacua 
is very flat. The tunneling in such cases can be described by the Hawking-Moss 
instanton, which is an Euclideanized dS space with the field at the top of the 
barrier. The corresponding transition rate is \cite{Hawking:1981fz}
\beq
\kappa_{HM} \sim e^{S_b-S_p},   
\label{HM}
\eeq
where $S_b$ is the entropy of the unstable dS vacuum at the top of the barrier.  This 
rate can also be derived \cite{Linde:1991sk} using the quantum diffusion formalism 
of eternal inflation.  Our focus in this paper will be on relatively steep potentials, such 
that the classical dynamics dominates and quantum diffusion can be neglected.

For upward and downward transitions between two dS vacua, the instanton 
solution is the same and the transition rates are related by  
\beq
\frac{\kappa_\uparrow}{\kappa_\downarrow} = e^{-\Delta S},
\label{kappaupdown}
\eeq
where $\Delta S$ is the entropy difference between the lower and higher dS 
vacua.  This relation can be interpreted as an expression of detailed balance 
between dS transitions in the multiverse.  It fits very well with the widely accepted 
picture of quantum dS space as a thermal state \cite{Dyson:2002pf}. It should 
be noted, however, that while Coleman's flat space calculation was solidly based 
on first principles, the CdL formula (\ref{CdL}) was proposed in \cite{Coleman:1980aw} 
essentially by analogy with the flat space case, so its validity is open to question.     

In this paper we shall use the stochastic approach to estimate vacuum decay rate 
in various settings, both with and without gravity.  We shall adopt the version of this 
approach introduced by Brown and Dahlen in Ref.~\cite{Brown:2011ry}.  Instead of an
initial fluctuation of the field $\phi$, they considered a fluctuation of the field velocity 
${\dot\phi}$ with $\phi$ remaining at its parent vacuum value.  We find this preferable, 
since representing the fluctuations by a free quantum field is better justified in this case, 
but we have verified that using initial field fluctuation with ${\dot\phi}=0$ would yield similar 
results. If the fluctuation is large enough and occurs in a large enough region, the field 
will "fly over" the barrier and form an expanding bubble of the daughter 
vacuum. We shall refer to such vacuum transitions as flyover transitions.

In the next section we estimate the rate of flyover transitions in flat space, considering 
the limits of thin-wall and thick-wall bubbles.  In both cases we find agreement with the 
instanton approach, up to ${\cal O}(1)$ factors in the exponent.
We also performed numerical simulations of flyover transitions.  They suggest that bubble 
formation and early evolution in these transitions can be rather different from the standard 
picture.

Flyover decay of dS vacua is discussed in Section III.  For the downward transition rate, once 
again we find agreement with the instanton formalism.  For upward transitions, however, the 
flyover rate is parametrically different and can be many orders of magnitude higher than 
suggested by the instanton calculation.  In contrast to the instanton approach, the bubble 
size in this case can be much smaller than the parent horizon.  The inflating false vacuum 
bubble is then contained inside of a black hole.

Our conclusions are summarized and discussed in Section IV.  Some technical details are 
given in the Appendices.

\section{Flat spacetime}

We first consider flyover vacuum decay in flat spacetime, ignoring the effects of 
gravity.  In this case upward transitions are forbidden by energy conservation, so we only 
consider downward transitions.

\subsection{Analytic estimates}
\begin{figure}
   \centering
\includegraphics[scale=0.25]{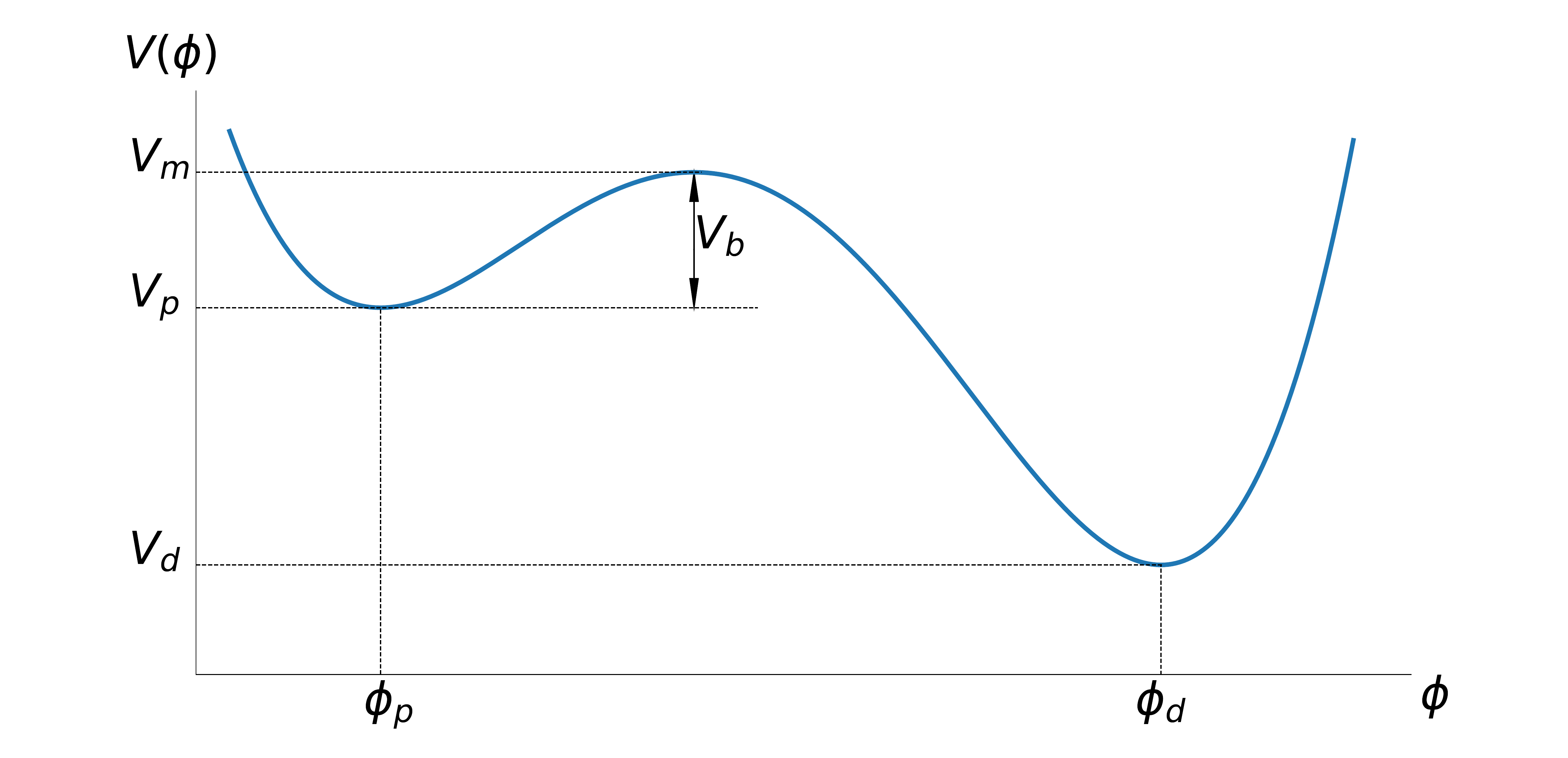}
\caption{\label{po1}{Potential with a true vacuum $V_d$ and a false vacuum $V_p$.}}
\label{genericpotential}
\end{figure}
Consider a scalar field $\phi$ with a double well potential of the form illustrated in Fig. \ref{po1}. 
Initially $\phi$ is in the parent vacuum, $\phi_p=0$. Suppose that at some time $t_0$ the field 
$\phi$ developed a velocity fluctuation with a spherically symmetric Gaussian profile,
\beq
{\dot\phi}(t_0,r)= {\dot\phi}_0 e^{-\frac{r^2}{2l^2}},
\label{fluc}
\eeq
with $\phi(t_0,r)=\phi_p$.  In order for the field to fly over the barrier, we need 
${\dot\phi}^2 \gtrsim 2V_b$, where $V_b=V_m - V_p$ is the height of the barrier, 
$V_m$ is the potential at the top of the barrier and $V_p$ is the potential in the parent 
vacuum.  In order for this condition to be satisfied for $r\lesssim l$, we need 
${\dot\phi}_0^2 \gtrsim 2eV_b$.

For $\phi$ near $\phi_p$ the potential has the form $V(\phi)\approx m^2\phi^2 /2$, 
where $m^2=V''(\phi_p)$ is the mass parameter of the field in the parent vacuum.  For 
a generic potential we expect $V_b \sim m^2 \Delta\phi^2$, where $\Delta\phi$ is the 
width of the barrier.  The bubble wall thickness $\delta$ is determined by the balance 
between the gradient and potential energies; it can be estimated by minimizing 
$[(\Delta\phi/\delta)^2 +V_b] \delta$.  This gives $\delta\sim \Delta\phi/\sqrt{V_b} \sim 1/m$.  
The wall tension is then $\sigma\sim V_b \delta \sim V_b/m$.

The force of tension acting per unit area of the bubble wall is $\sigma/R$, and the 
force due to  the difference of vacuum energy densities is $V_p-V_d\equiv\epsilon$, 
where $R$ is the bubble radius and $V_p$ and $V_d$ are the energy densities of 
the parent and daughter vacua, respectively.  For a bubble to expand, its radius 
should exceed $2\sigma /\epsilon\equiv R_0$.\footnote{The initial bubble radius 
given by analytic continuation of the Coleman's instanton is $3\sigma/\epsilon$, but 
the smallest bubble that can expand has radius $2\sigma/\epsilon$.}  A thin-wall 
bubble with $R_0\gg\delta$ corresponds to $V_b \gg\epsilon$.  The opposite limit, 
where $V_b\ll\epsilon$ and $R_0\ll\delta$ will be referred to as a thick-wall bubble.

For a free scalar field the fluctuations are Gaussian and the probability of a velocity 
fluctuation of magnitude ${\dot\phi}_0$ on scale $l$ is \cite{Huang}
\beq
P \sim \exp\left(-\frac{{\dot\phi}_0^2}{2\langle {\dot\phi}_l^2\rangle} \right) ,
\label{Pflat}
\eeq
where ${\dot\phi}_l$ is the field velocity operator smoothed with a Gaussian window 
of radius $l$ and the angular brackets indicate a vacuum expectation value. Note 
that this is true due to the fact that the smoothing process preserves the Gaussian 
nature of the smeared fluctuations.  This expectation value is calculated in 
Appendix A, with the result
\beq
\langle{\dot\phi}_l^2\rangle \approx \frac{1}{8\pi^2 l^4}
\label{rms1}
\eeq
for $ml\ll 1$, and
\beq
\langle{\dot\phi}^2_l\rangle \approx \frac{m}{16\pi^{3/2} l^3}
\label{rms2}
\eeq
for $ml\gg 1$.  
In our case $\phi$ can be accurately approximated as a free field only for $\phi\ll\Delta\phi$, 
but we shall assume that the above estimates are also correct by order of magnitude 
for $\phi\sim \Delta\phi$.  

In the thin-wall case, the minimal length scale of a fluctuation that can result in an 
expanding bubble is $l \sim \sigma/\epsilon$.  Then $ml\sim V_b/\epsilon\gg 1$ and 
we obtain the following estimate for the transition rate:
\beq
\kappa \sim \exp\left(-16e\pi^{3/2}\frac{V_b l^3}{m}\right).
\label{flatrate}
\eeq
Using $V_b/m\sim \sigma$ and $l\sim \sigma/\epsilon$, this can be rewritten as
\beq
\kappa \sim \exp\left(-16e\pi^{3/2}\frac{\sigma^4}{\epsilon^3}\right).
\label{flatratethin}
\eeq
In the following subsection we will show a numerical example
in this thin-wall limit, where this flyover transition will happen if we take
$l \sim 4 \sigma/\epsilon$. Apart from the numerical coefficient in the exponent, this 
agrees with Coleman's result
\beq
\kappa_{\rm Coleman}\sim \exp\left(-\frac{27\pi^2}{2}\frac{\sigma^4}{\epsilon^3}\right).
\eeq	
In the thick-wall case, the potential in the vicinity of the barrier can be approximated as
\beq
V(\phi)\approx \frac{1}{2}m^2\phi^2 -\gamma \phi^3
\eeq
and the tunneling action is \cite{Sarid:1998sn,Dine:2015ioa} 
\beq
S\approx 24\frac{m^2}{\gamma^2}.
\label{Dine}
\eeq
The height of the barrier in this case is $V_b = m^6/54\gamma^2$ and the minimal fluctuation 
size is $l\sim \delta\sim 1/m$. More accurately, numerical simulations described in the following 
subsection indicate that $l \sim 3/m$.  Substituting this in  the exponent of Eq.~(\ref{flatrate}) gives  
\beq
\ln\kappa\sim 10^2 \frac{m^2}{\gamma^2} .
\label{flatratethick}
\eeq
This has the same dependence on the parameters as in (\ref{Dine}), but the numerical coefficient is several times larger.

The difference in the numerical coefficient is not surprising if we note that {\it (i)} we 
estimated the length scale $l$ only by order of magnitude and that {\it (ii)} we used a 
free field estimate for the rms velocity fluctuation.  Furthermore, we assumed a specific 
form of the initial fluctuation (with velocity fluctuating and the field value remaining fixed), 
which is not necessarily the optimal fluctuation resulting in a bubble.  

In the above analysis we assumed that the fluctuation keeps its coherence during the 
transition and obeys the classical field equation.\footnote{See also 
Ref.~\cite{Hertzberg:2019wgx} for a recent discussion of this issue.}  The fluctuation 
is localized in a region of size $\sim l$ and is composed of the field modes of 
wavelength $\sim l$ and momentum $k\sim 1/l$.   These modes will disperse on a 
time scale $\tau \sim l/v \sim ml^2$, where $v\sim k/m$ is the characteristic 
velocity.  For the fluctuation to remain coherent it is necessary that $\tau\gtrsim \Delta t$, 
where $\Delta t \sim \Delta\phi /{\dot\phi} \sim 1/m$ is the flyover time.\footnote{We are 
grateful to Ken Olum for suggesting this estimate.}  This requires that $ml\gtrsim 1$, 
which is satisfied with a great margin for a thin-wall bubble and is marginally satisfied 
for a thick-wall bubble.  The fluctuation could also disintegrate due to the fact that its 
constituent modes would oscillate out of phase.  But the oscillation frequency is 
$\sim m$ and once again the dispersion time is $\tau\sim 1/m$.

We note also that bubble nucleation in flat spacetime should conserve energy.  On 
the other hand, the velocity fluctuation (\ref{fluc}) has a positive energy 
${\cal E}=(2\pi)^{3/2}{\dot\phi}_0^2 l^3$.  This energy should be 
compensated by a negative energy fluctuation of the vacuum.  
The mechanism of this energy compensation and its effect on bubble 
formation require further analysis.

A similar approach can be applied to bubble nucleation at a finite temperature $T$.  It is 
shown in the Appendix that the rms field velocity fluctuation in this case can be represented 
as a sum of vacuum and thermal contributions,
\beq
\langle{\dot\phi}_l^2\rangle = \langle{\dot\phi}_l^2\rangle_v +\langle{\dot\phi}_l^2\rangle_T ,
\eeq
where the vacuum contribution is the same as that at $T=0$.  At $T\ll m$ the thermal 
contribution is exponentially suppressed and the nucleation rate is nearly the same as 
in vacuum.  On the other hand, at $T\gg m, 1/l$ thermal fluctuations dominate and we 
find
\beq
\langle{\dot\phi}_l^2\rangle_T  \approx \frac{T}{8\pi^{3/2} l^3}. 
\eeq
The nucleation rate is then
\beq
\kappa \sim \exp\left(-8e\pi^{3/2}\frac{V_b l^3}{T}\right) .
\label{thermal}
\eeq
Note that ${\cal E}\sim V_b l^3$ is the energy needed to create a bubble of size $l$ at 
the top of the barrier, so the right-hand side of (\ref{thermal}) can be interpreted as 
the Boltzmann suppression factor. Once again, the exponent in (\ref{thermal}) is 
parametrically the same as given by the instanton method \cite{Linde:1981zj}.

We note finally that the exponents in our estimates of flyover decay rates are consistently 
greater than those in the tunneling rates calculated by the instanton method, suggesting 
that flyover is a subdominant decay channel.  This conclusion, however, should be taken 
with caution, since the numerical coefficients in the flyover exponents are not expected 
to be accurate, for the reasons given above.

\subsection{Numerical simulations}

In this section, we will numerically investigate the possibility of obtaining an expanding
region of true vacuum by the "flyover" process we discussed earlier. In order to do this 
we will consider a scalar field with a potential of the form
\beq
V(\phi) = \frac{1}{2} m^2 \phi ^2 - \gamma \phi^3 + \lambda  \phi^4,
\label{Sarid-po}
\eeq
where we will vary the parameters in such a way that we will always enforce the
presence of two non-degenerate vacua separated by a potential barrier. We will
not be concerned with the value of the cosmological constant in each vacuum since in 
this case we will consider this process in flat space.

As we mentioned earlier we will consider the situation where the initial fluctuation
is of the form
\beq
{\dot\phi}(t_0,r)= {\dot\phi}_0 e^{-\frac{r^2}{2l^2}},
\eeq
where $l$ describes the typical scale of the initial fluctuation and ${\dot\phi}_0$
denotes the maximum fluctuation amplitude of this initial state. Assuming spherical 
symmetry we can evolve this initial state by solving the equation
\beq
\ddot \phi -  \phi''  - \frac{2}{r} ~\phi' + \frac{\partial V (\phi)}{\partial \phi} = 0 ~.
\eeq
where we impose the boundary condition at the center of the fluctuation to be such that
\beq
\phi' (r=0) =0
\eeq
and we take the outer boundary of the simulation box to be fixed at the false vacuum 
value of the field. 

Following Ref. \cite{Sarid:1998sn} we can always rescale the field and the coordinates 
by the following redefinitions, $\tilde \phi = (2\gamma/m^2) \phi$,
$\tilde x^{\mu} = (m/\sqrt{2}) x^{\mu}$ to obtain a potential that 
depends only on the dimensionless quantity, $\iota=1- 2 \lambda m^2/\gamma^2$,
namely,
\beq
\label{potential-parametrization}
\tilde V(\tilde \phi) = \tilde \phi ^2 - \tilde \phi^3 +  \frac{1}{4} (1- \iota) \tilde \phi^4,
\eeq
where $\tilde{V}=(8\gamma^2/m^6)V$. In the following we will investigate two scenarios
with different values of $\iota$.

\subsubsection{Thin-wall case}

Taking the limit of $\iota \rightarrow 0$ the two minima of the potential become 
degenerate. This corresponds to the thin-wall limit of the usual instanton calculation.
The top of the potential is in this case $ \tilde V_b \approx 1/4$, the potential difference 
in dimensionless units is given by $\tilde \epsilon \approx 4 \iota$ and the interpolating wall
has a tension\footnote{This tension can be easily calculated 
using the kink solution interpolating between the two degenerate vacua obtained by setting 
$\iota=0$, namely $\tilde \phi_K(\tilde x) = \tanh (\tilde x/\sqrt{2})$. The tension in this case can 
be found to be $\tilde \sigma \approx \int{\left(\frac{1}{2} \tilde\phi'^2 + \tilde V(\tilde \phi) \right) d\tilde x}\approx 2\sqrt{2}/3$.} $\tilde \sigma  \approx 2\sqrt{2}/3$.

As we described in the previous section, we are interested in studying a fluctuation
with a central amplitude given $\dot{\tilde{\phi_0}} \approx \left(2 e \tilde V_b\right)^{1/2}\approx  \sqrt{e/2}$ 
and whose scale is of the order of $\tilde l \approx C \tilde \sigma/\tilde \epsilon \approx C \sqrt{2}/(6\iota)$, 
where $C$ is a numerical coefficient of the order one that we will find from our simulations.

\begin{figure}
   \centering
\includegraphics[scale=0.25]{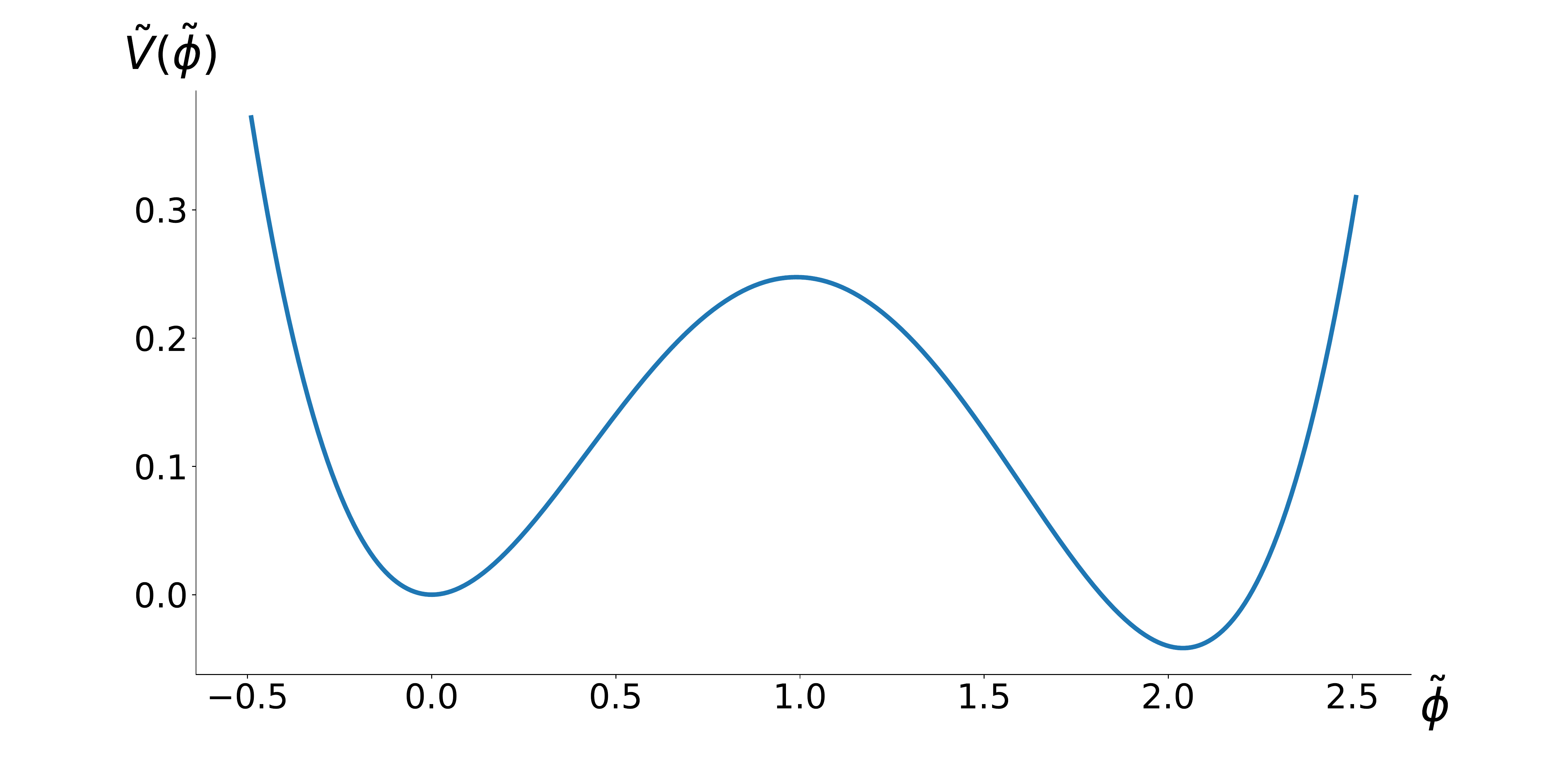}
\caption{\label{thin_po}{Plot of the potential in Eq. (\ref{potential-parametrization}) with $\iota = 0.01$.}}
\end{figure}

\begin{figure}
\includegraphics[scale=0.5]{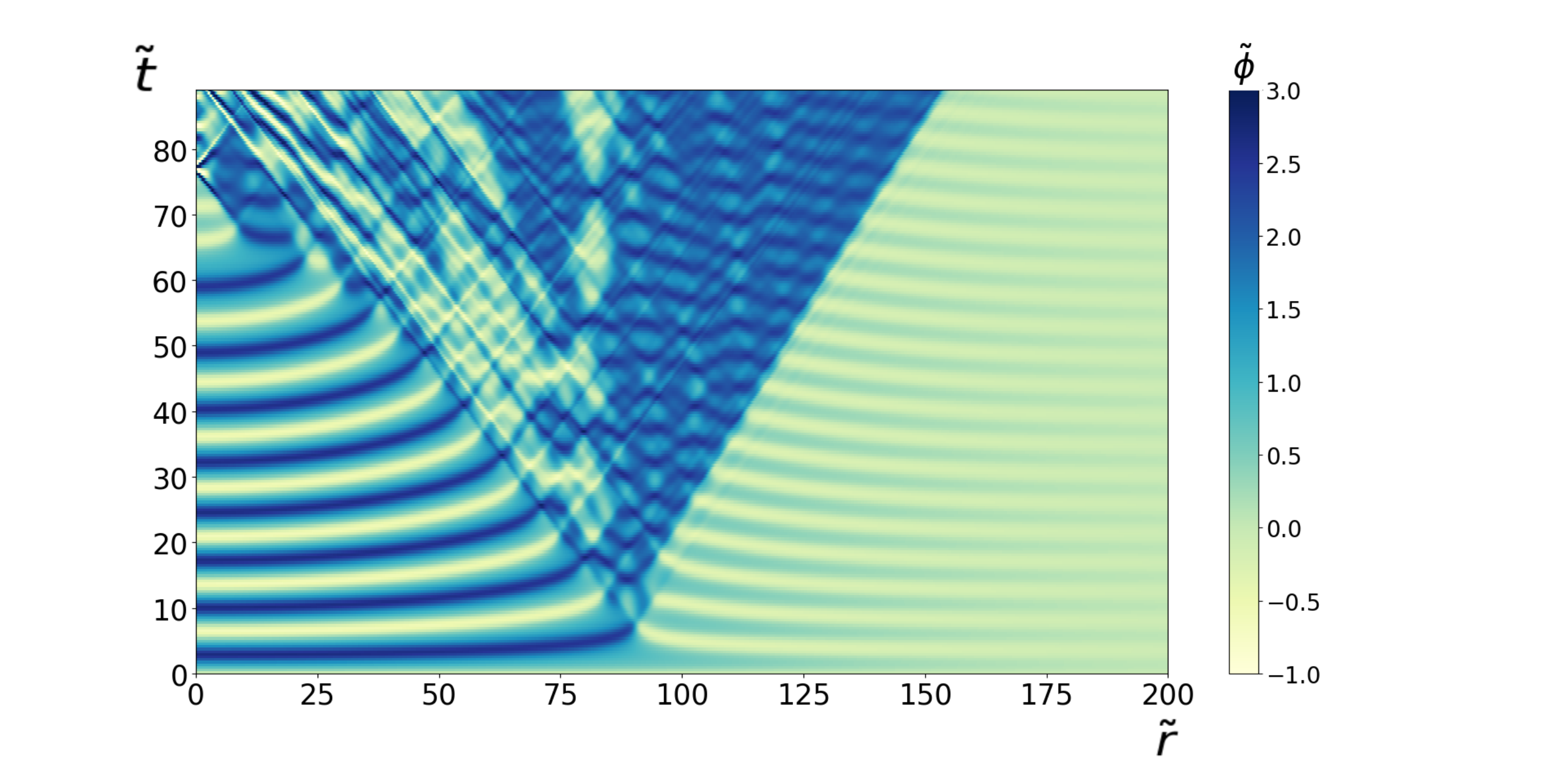}

\caption{Evolution of the scalar field 
under the potential with $\iota=0.01$ ($\tilde{\phi}_d\approx 2$). The initial fluctuation size 
is $\tilde{l}=4 \tilde{\sigma} / \tilde{\epsilon}\approx 90$. A shell is formed at $\tilde r\approx90$, 
with the outer boundary propagating outwards and the inner inwards. Both boundaries propagate 
at a speed close to that of light.
}
\label{fig:thin-wall-flat}
\end{figure}

We show in Fig. \ref{fig:thin-wall-flat} an example of the evolution of this kind of initial configuration 
with $\iota=0.01$ (the potential is shown in Fig. \ref{thin_po}) and 
$\tilde{l}=4 \tilde{\sigma} / \tilde{\epsilon}$. We can see in the figure that the central 
region of the fluctuation quickly evolves over the barrier. However, 
this region is too small to create an expanding bubble. This leads 
to an oscillating pattern between the false and the true vacuum near the center of the fluctuation.

Further away from the center, the true vacuum is reached later on in
the simulation when the interior has already bounced back to the false vacuum. 
We found that, when the initial size of the fluctuation is large enough, a shell of radius $\sim \tilde l$ 
will form that interpolates between oscillating interior and the parent vacuum outside.
The subsequent evolution of this shell is such that its inner boundary propagates
inwards and the outer boundary outwards. Both boundaries propagate at a speed 
close to that of light.  For $\iota=0.01$ the smallest $\tilde l$ that is able to produce such a shell 
was found to be $\tilde l \approx 4\tilde{\sigma} / \tilde{\epsilon}$. 

The field in the expanding shell oscillates around the true vacuum, while in the central region 
inside of the inner boundary of the shell, it oscillates between the true and the false vacuum.

\subsubsection{Thick-wall case}
We now study the other end of the spectrum of the possible values of the parameters where the
instanton calculation would lead to the formation of a thick bubble. In our parametrization this is 
achieved by taking the limit of $\iota\rightarrow 1$. 

\begin{figure}
   \centering
\includegraphics[scale=0.25]{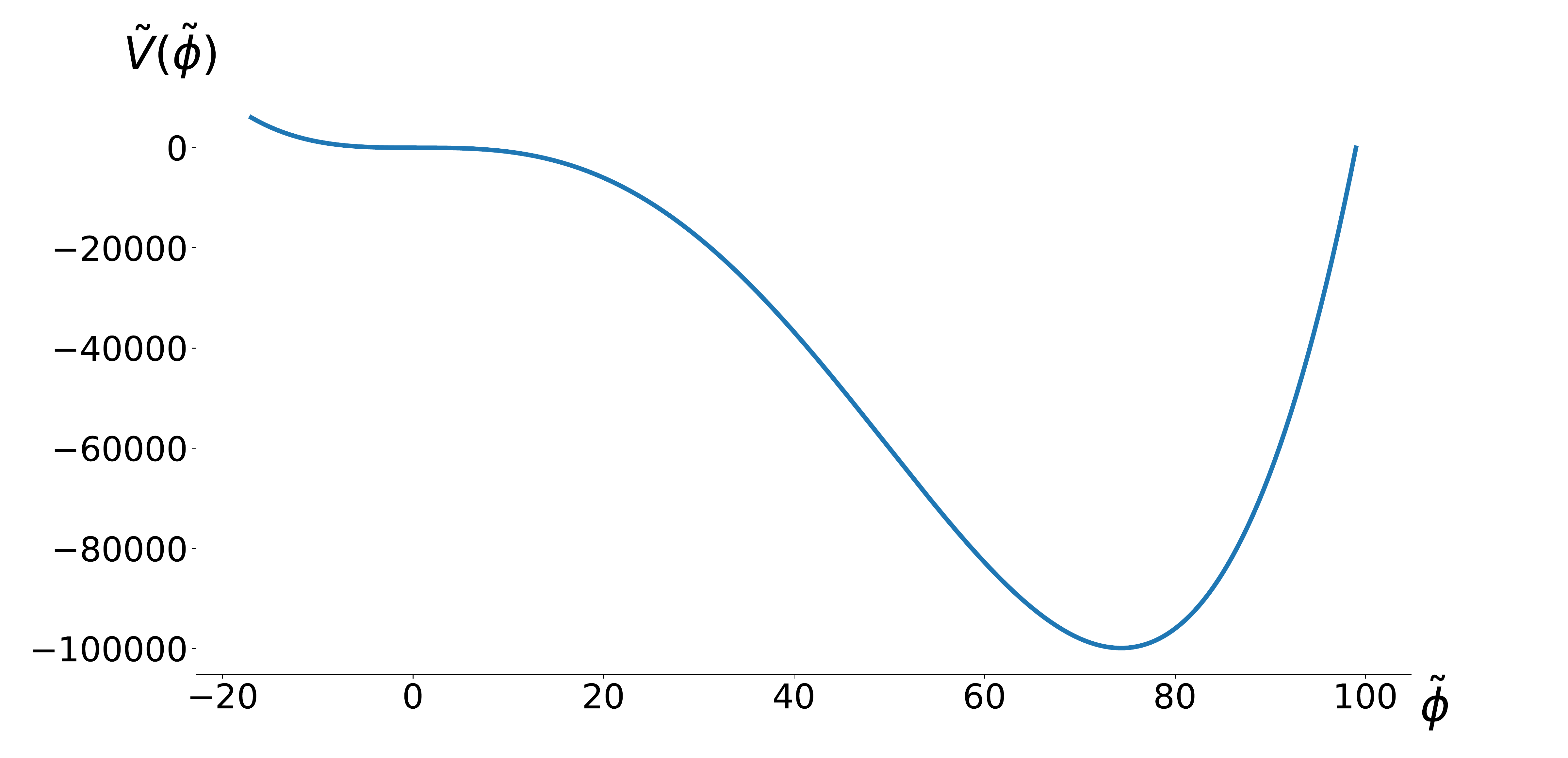}
\caption{\label{thick_po}{Potential with $\iota = 0.96$.}}
\end{figure}

Taking the form of the potential with $\iota=0.96$ (see Fig. \ref{thick_po}) we have
found that with $\dot{\tilde{\phi_0}} = \sqrt{2 e \tilde V_b}$, the smallest fluctuation size that is able 
to produce a bubble is $\tilde l \approx 2.2$. This gives $ml \approx 3$, which leads to our estimate in Eq. (\ref{flatratethick}).

\begin{figure}
\includegraphics[scale=0.5]{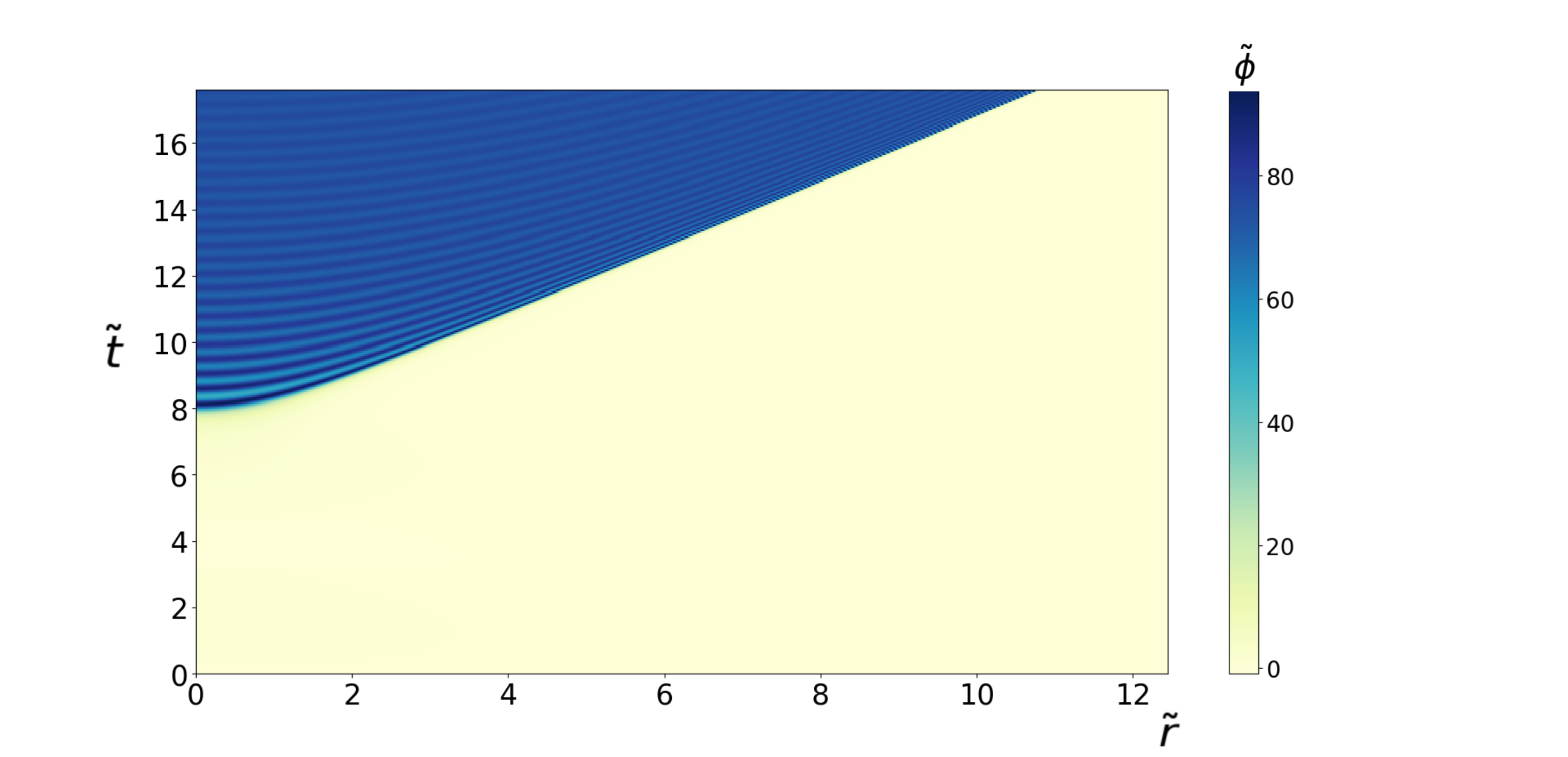}
\caption{Evolution of the scalar field under the potential with $\iota=0.96$ 
($\tilde{\phi}_d\approx 74$). The initial fluctuation size is $\tilde{l}=2.2$.}
\label{thick2}
\end{figure}

We see in Fig. \ref{thick2} how, in this case, the initial fluctuation 
brings a large enough region of space over the barrier, so the resulting bubble of true
vacuum is able to expand similarly to what happens in the usual instanton decay.

Since the time it takes for the field at the center to reach the true vacuum is of the 
order $\Delta \phi / \dot{\phi}_0 \sim 1/m$ which in this case is $\sim l$, almost the 
whole fluctuation region gets to the true vacuum simultaneously. As we can see from Fig. \ref{thick2}, the shell structure in the thin-wall 
case completely disappears, and the field inside the bubble oscillates around the true vacuum.

As a final remark, let us note that rescaling things back we can get a family of solutions 
with the same kind of behavior as the one given in any of the scenarios presented here 
but written in terms of the original physical parameters, $m, \gamma$ and $\lambda$.

\section{Inflating universe}

We now turn to flyover transitions in an inflating universe.  The potential we consider 
still has the form shown in Fig.~(\ref{genericpotential}).
We further assume the height of the barrier to be small compared to the vacuum energy 
in the parent vacuum, $V_b\ll V_p$.  This ensures that the gravitational back-reaction of 
the velocity fluctuation is insignificant.   
Also, in the absence of fine-tuning we expect the mass of the field in the parent vacuum to satisfy $m\gg H_p$.
Both parent and daughter vacua are de Sitter, so transitions can occur both upwards and downwards.

\subsection{Downward transitions}

We first consider downward transitions in the case of a small bubble, when the bubble 
size is small compared to the parent horizon $H_p^{-1}$.  For a thick-wall bubble this follows 
from the condition $m\gg H_p$, while for a thin-wall bubble this requires $l_0\ll H_p^{-1}$, 
where $l_0 \sim \sigma/\epsilon$ is the flat space bubble size.   The minimal velocity necessary 
to take the field over the barrier is ${\dot\phi}_0^2\sim 2C V_b$, where the factor $C > 1$ 
accounts for the Hubble friction.  The flyover time is $\Delta t\sim \Delta\phi/{\dot\phi}_0 \sim 1/m \ll 1/H_p$.  
Since it is small compared to the Hubble time, we expect Hubble friction to be insignificant and $C\sim 1$.
The rms velocity fluctuation $\langle {\dot\phi}^2_l\rangle$ in dS space is estimated in 
Appendix A.  Somewhat surprisingly, it is given by the same limiting expressions 
(\ref{rms1}), (\ref{rms2}) as in flat space, independent of the expansion rate $H_p$.  Also, 
for a small bubble, the minimal length scale of the fluctuation needs to be the same as in 
flat space, that is, $l\sim l_0$ and $l\sim 1/m$ for thin-wall and thick-wall bubbles, 
respectively.  The bottom line is that our estimates for the rate of downward transitions 
are still given by Eqs.~(\ref{flatrate}) and (\ref{flatratethick}), the same as in flat space.

Let us now consider a downward transition with a large bubble, whose size is not small 
compared to the horizon.  This has to be a thin-wall bubble, since the wall thickness is 
$\delta\sim 1/m\ll H_p^{-1}$.  Hence we must have $\epsilon\ll V_b \ll V_p$, so the energy 
densities of the parent and daughter vacua should be very close to one another: $V_p\approx V_d$.
The minimal size of an inflating bubble in this case is $l\sim H_p^{-1}$.  Substituting this 
in Eq.~(\ref{flatrate}) we find 
\beq
\kappa\sim \exp\left(-16 e\pi^{3/2}\frac{V_b}{mH_p^3}\right) \sim \exp\left(-16e\pi^{3/2}\sigma H_p^{-3}\right).
\label{dSrate}
\eeq
This can be compared with the instanton calculation of the nucleation rate of domain 
walls in dS space \cite{Basu:1991ig}
\beq
\kappa_{\rm wall} \sim \exp\left(-2\pi^2 \sigma H_p^{-3}\right).
\eeq
(Our process is similar to domain wall nucleation, since $V_p\approx V_d$.).
Once again the two expressions agree, up to a numerical factor in the exponent.

\subsection{Upward transitions}

\begin{figure}
   \centering
\includegraphics[scale=0.25]{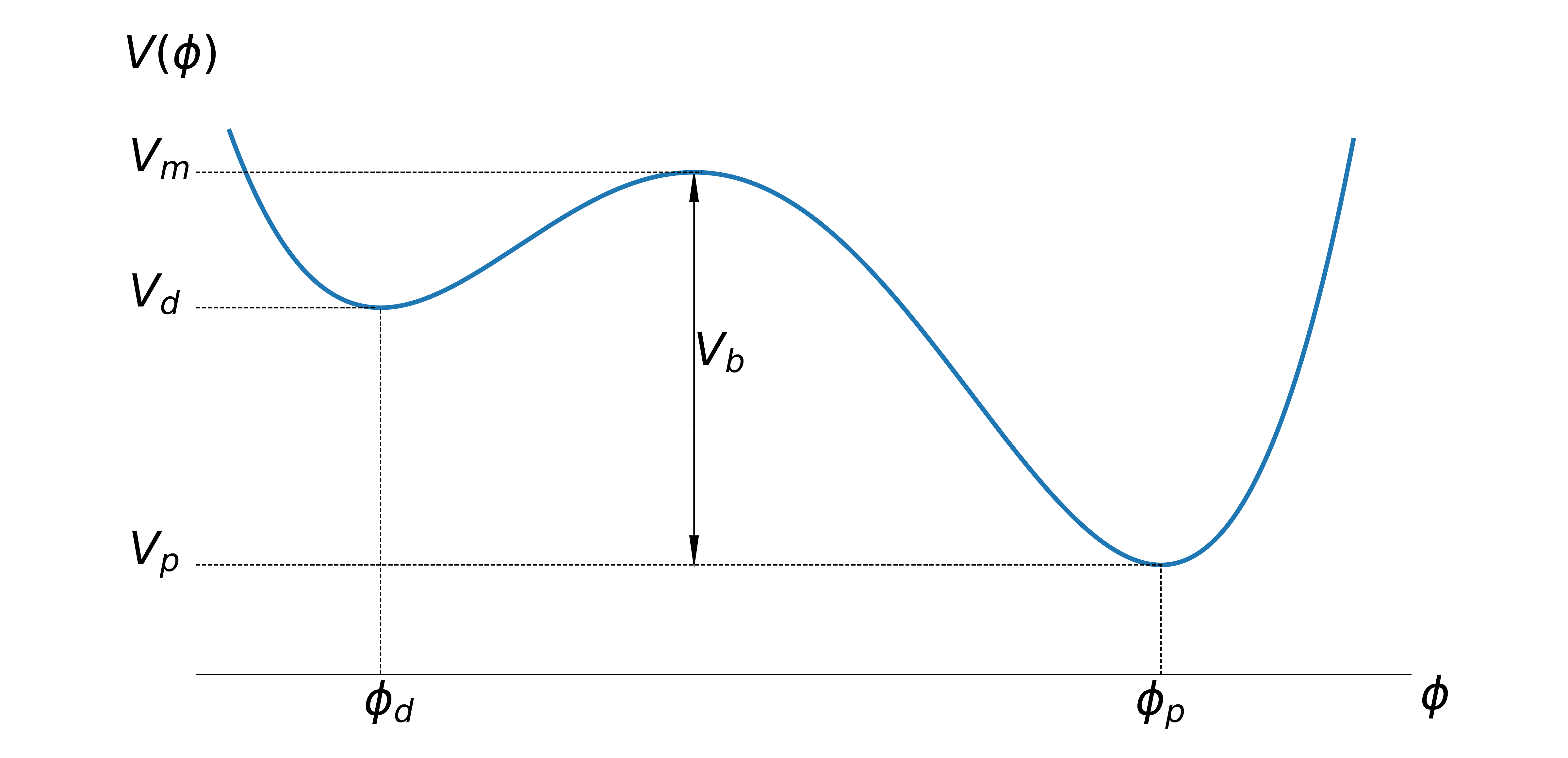}
\caption{\label{po2}{Potential with a true vacuum $V_p$ and a false vacuum $V_d$.}}
\end{figure}

For an upward transition (Fig. \ref{po2}), we have $\epsilon\lesssim V_b \ll V_p$; hence 
we should have $V_p\approx V_d$ and $H_p\approx H_d$.  With either relation between 
$l_0$ and $H_p^{-1}$, the bubble will inflate only if its size is greater than the horizon, 
$l\gtrsim H_p^{-1}$.  The transition rate is then given by Eq.~(\ref{dSrate}).  This has 
some interesting implications.  

The upward and downward transition rates can generally be expressed as 
$\kappa_\uparrow \sim e^{-B_\uparrow}$ and $\kappa_\downarrow \sim e^{-B_\downarrow}$, 
with $B_\uparrow \sim \sigma H_p^{-3}$ and $B_\downarrow \sim \sigma l^3$, where 
$l = {\rm min} \{l_0,H_p^{-1} \}$.  The ratio of these rates is
\beq
\frac{\kappa_\uparrow}{\kappa_\downarrow} = e^{-\Delta B} ,
\label{ratio}
\eeq
where 
\beq
\Delta B = B_\uparrow - B_\downarrow \lesssim \sigma H_p^{-3}.
\label{DeltaB}
\eeq
According to the instanton calculation, this ratio must satisfy the detailed balance 
condition (\ref{kappaupdown}), which requires that $\Delta B = \Delta S$, where 
\beq
\Delta S = S_p - S_d \sim \frac{\epsilon}{H_p^4}
\label{DeltaS}
\eeq
is the difference of the de Sitter entropies of the two vacua.  Now, from Eqs.~(\ref{DeltaB}) 
and (\ref{DeltaS}) we can write
\beq
\Delta B \lesssim \frac{l_0}{H_p^{-1}} \Delta S.
\eeq
This can be consistent with detailed balance if $l_0 \gtrsim H_p^{-1}$.  However, for 
$l_0 \ll H_p^{-1}$ we have $\Delta B \ll \Delta S$ and the detailed balance is strongly 
violated.  In this case the upward flyover transition rate is much greater than predicted 
by the instanton formalism.

To be more specific, consider a model where
\beq
V_p - V_d  \sim  V_b  \ll  V_p  \ll {V_b}^{1/2}.
\label{*}
\eeq
Let us also assume that the barrier is characterized by a single mass 
scale $m$. Then the flat-space version of this model describes a 
thick-wall bubble of initial size $l_0 \sim m^{-1} \sim V_b^{-1/4}$, and 
the last inequality in (\ref{*}) ensures that $l_0 \ll H_p^{-1}$, where 
$H_p \sim V_b^{1/2}$.  This implies 
that the bubble size and the transition amplitude for downward transitions 
are unaffected by gravity; in particular the bubble size $l$ is close to 
the flat space value $l_0$.  The condition of small gravitational 
back-reaction is also satisfied, and thus we have a well controlled model 
where detailed balance is strongly violated.

As we already indicated, our analysis cannot be directly extended to large upward 
jumps with $V_d\gg  V_p$. In this case an inflating bubble can be formed by a fluctuation 
of size $l\sim H_d^{-1}$, which can be much smaller than the parent horizon $H_p^{-1}$.  This 
bubble will inflate at a rate much higher than that of the parent vacuum, which is possible 
geometrically only if the daughter bubble is connected to the parent vacuum by a wormhole.  
The situation here is similar to baby universe formation from false vacuum bubbles produced 
during inflation, as discussed in Refs.~\cite{Garriga:2015fdk,Deng:2017uwc}.  It is also similar 
to the formation of an inflating baby universe in a laboratory, as discussed by Farhi, Guth and 
Guven \cite{Farhi:1989yr} (See also \cite{Garriga:2004nm,Aguirre:2005nt}). The wormhole must close off in about a light crossing time, 
resulting in an inflating daughter region contained in a baby universe inside a black hole.  The 
mass of the black hole can be estimated as
\beq
M\sim V_d H_d^{-3} \sim H_d^{-1}.
\label{MBH}
\eeq
This scenario is confirmed by a numerical simulation described in the next subsection.

We cannot reliably estimate the rate of large upward transitions, since such transitions 
would induce a large gravitational back-reaction.  As a plausible guess, one can expect 
that it is comparable to the nucleation rate of black holes of mass (\ref{MBH}) in the parent dS space:
\beq
\kappa \sim \exp\left(-\frac{M}{T}\right) \sim \exp\left(-\frac{C}{H_d H_p}\right) .
\eeq
Here $T=H_p/2\pi$ is the Gibbons-Hawking temperature of the parent vacuum and $C$ is a 
numerical coefficient ${\cal O}(1)$. Once again, this rate is much higher than the estimate 
based on the Lee-Weinberg instanton, 
\beq
\kappa_{LW}\sim \exp\left(-\frac{\pi}{H_p^2}\right).
\eeq

\subsection{Numerical simulations of upward transitions}

We will now present some numerical results of the upward transitions discussed in the last 
subsection, where the background is inflating and gravity can no longer be neglected. We 
solve the scalar field equation together with Einstein's equations for the spherically symmetric
initial fluctuation of the form given by Eq. (\ref{fluc}). 

The metric we consider is
\beq
ds^2=-dt^2+B^2 dr^2+R^2 d\Omega^2,
\eeq
where $d\Omega^2$ is the metric on a unit sphere and $B$ and $R$ (which is called the 
area radius) are functions of the coordinates $t$ and $r$. 
We give in Appendix B a detailed account of the numerical implementation of this system
of equations as well as the method we use to impose the initial conditions.

Specifically, we will use several examples to illustrate the flyover upward transitions in 
two cases: {\it (i)} $V_d \sim V_p \gg V_b$, where the background can be regarded as 
fixed dS; and {\it (ii)} $V_d \sim V_b \gg V_p$, where strong gravitational back-reaction 
is expected.

\subsubsection{$V_d \sim V_p \gg V_b$}
\begin{figure}
   \centering
\includegraphics[scale=0.25]{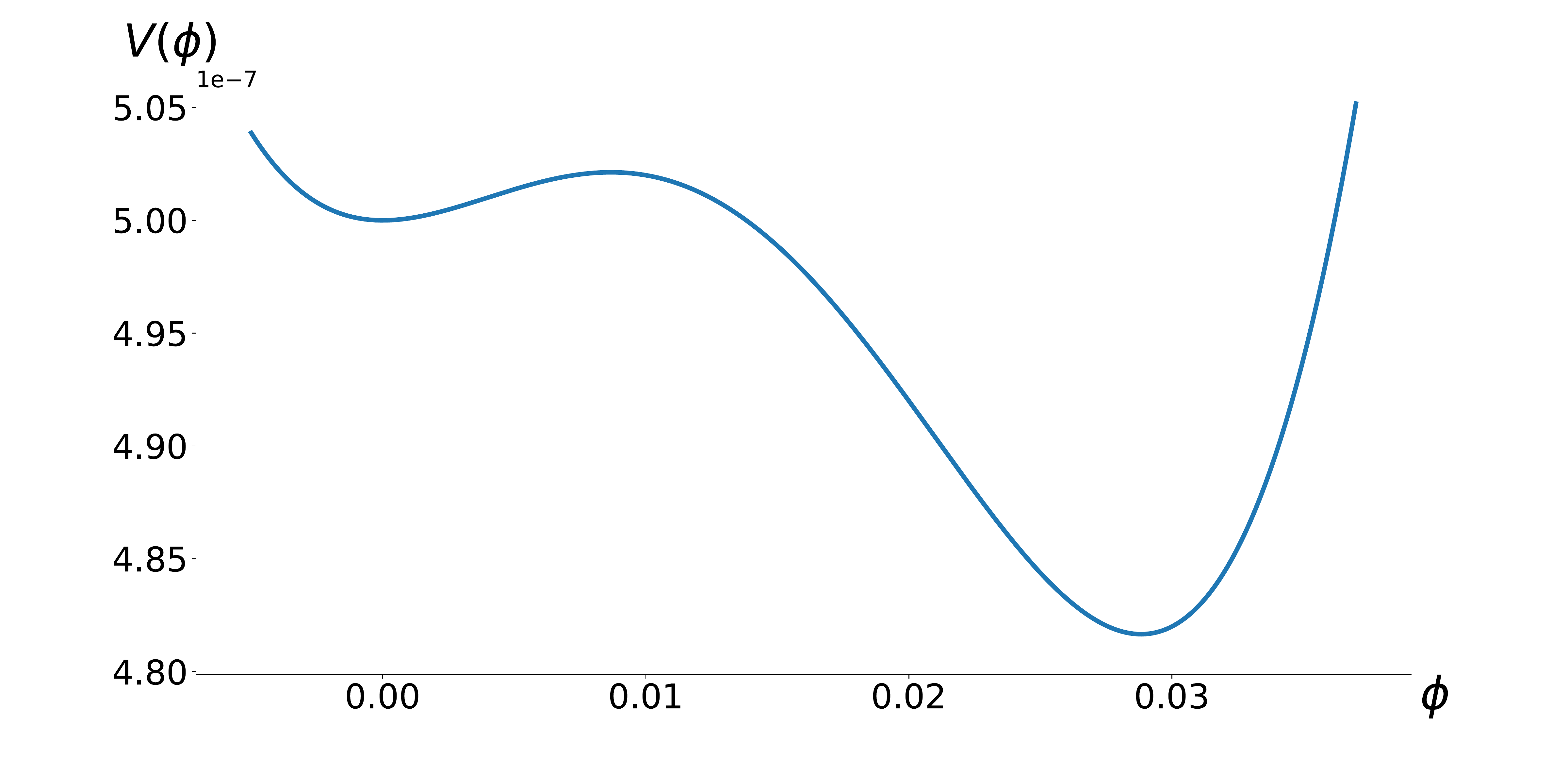}
\caption{\label{up-po-1}{Potential with $V_d \sim V_p \gg V_b$}}
\end{figure}

\begin{figure}
   \centering
\includegraphics[scale=0.5]{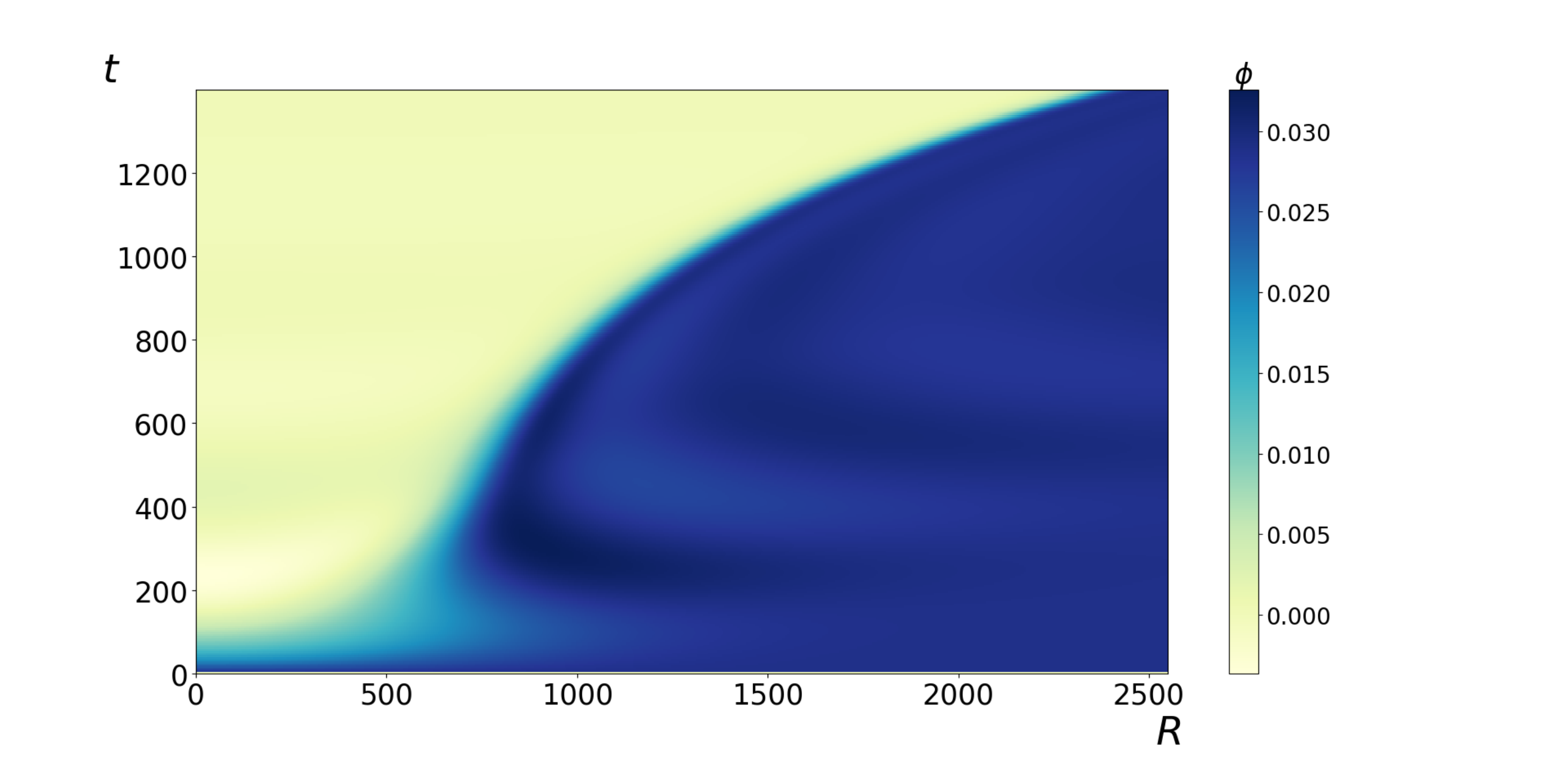}
\caption{\label{up-1}{Evolution of the scalar field under the potential Fig. \ref{up-po-1}, 
with $l=H_d^{-1}\approx 490$ and $\dot{\phi}_0 = -1.8\sqrt{2V_b}$. The horizontal axis 
is the area radius $R$ and the vertical axis is time $t$. We can clearly see the formation 
and growth of a false vacuum bubble.}}
\end{figure}

As before, we consider a scalar field with a potential in the form of Eq. (\ref{Sarid-po}),
\beq
 V(\phi) = 10^{-8} [(100\phi) ^2 - (100\phi)^3 + 0.2(100\phi)^4] + 5\times 10^{-7},
\eeq
where the parameters have been chosen such that we fullfill the requirements given above 
together with the condition $m \gg H_p$ while still being in a sub-Planckian regime.\footnote{In this section all quantities are in Planck units.}
In particular we get $m^{-1} \approx 46$, $l_0 \approx 225$, $H_p^{-1} \approx 500$ and 
$H_d^{-1} \approx 490$.  We show the form of this potential in Fig. \ref{up-po-1} where we 
can see that the energy difference between the two vacua is small compared to their absolute values.

\begin{figure}
   \centering
\includegraphics[scale=0.5]{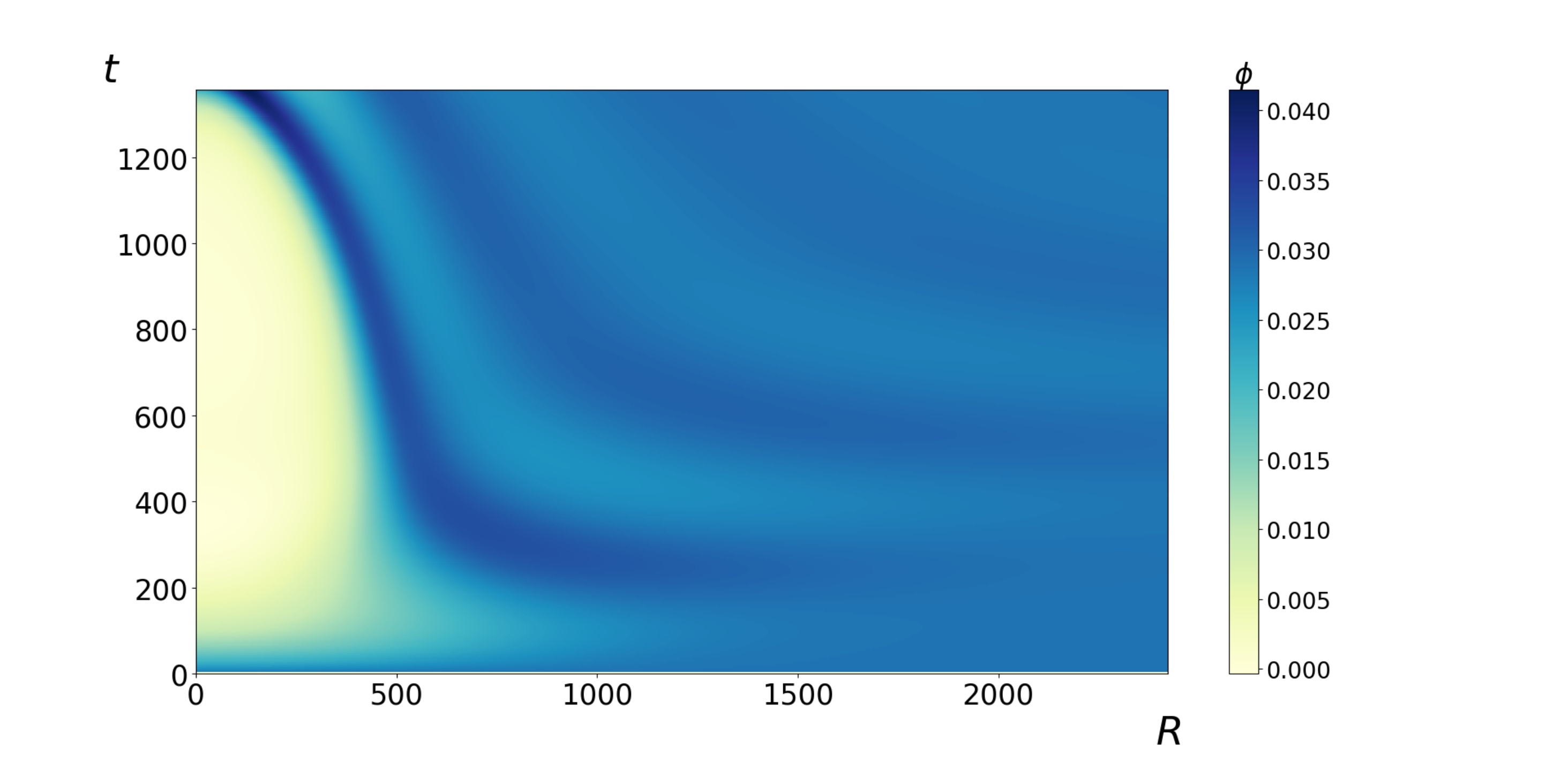}
\caption{\label{collapse1}{Evolution of the scalar field under the potential Fig. \ref{up-po-1}, 
with $l=H_d^{-1}\approx 490$ and $\dot{\phi}_0 = -1.6\sqrt{2V_b}$. The horizontal axis is 
the area radius $R$ and the vertical axis is time $t$. A false vacuum bubble forms and then 
collapses.}}
\end{figure}

We show in Fig. \ref{up-1} an example of the field evolution obtained using the equations
in the Appendix.  The field is initially in the true vacuum everywhere,
 $\phi(t_0,r)\approx 0.029$, but the velocity field has a spherically symmetric fluctuation 
with $l= H_d^{-1}$ and $\dot{\phi}_0 = -1.8\sqrt{2V_b}$. It can be seen that, as the field at the center of the 
fluctuation flies over the barrier and reaches the false vacuum valley, a bubble is formed 
and starts to grow exponentially. This confirms our theoretical expectations from the
previous section.

If the magnitude (or the size) of the field velocity fluctuation is not large enough, even if the
field near the center flies over the barrier and form a bubble, the bubble would eventually 
shrink and disappear. Fig. \ref{collapse1} shows such a case with $l= H_d^{-1}$ and 
$\dot{\phi}_0 = -1.6\sqrt{2V_b}$. The initial fluctuation has the same size as the one that 
gives Fig. \ref{up-1}, but a smaller magnitude. This case is nearly critical. As we 
can see from Fig. \ref{collapse1}, the bubble reaches a size close to the daughter horizon 
$H_d^{-1} \approx 490$, which means gravity almost defeats pressure and creates an 
inflating region. But the bubble ends up collapsing.

\subsubsection{$V_d \sim V_b \gg V_p$}
\begin{figure}
   \centering
\includegraphics[scale=0.25]{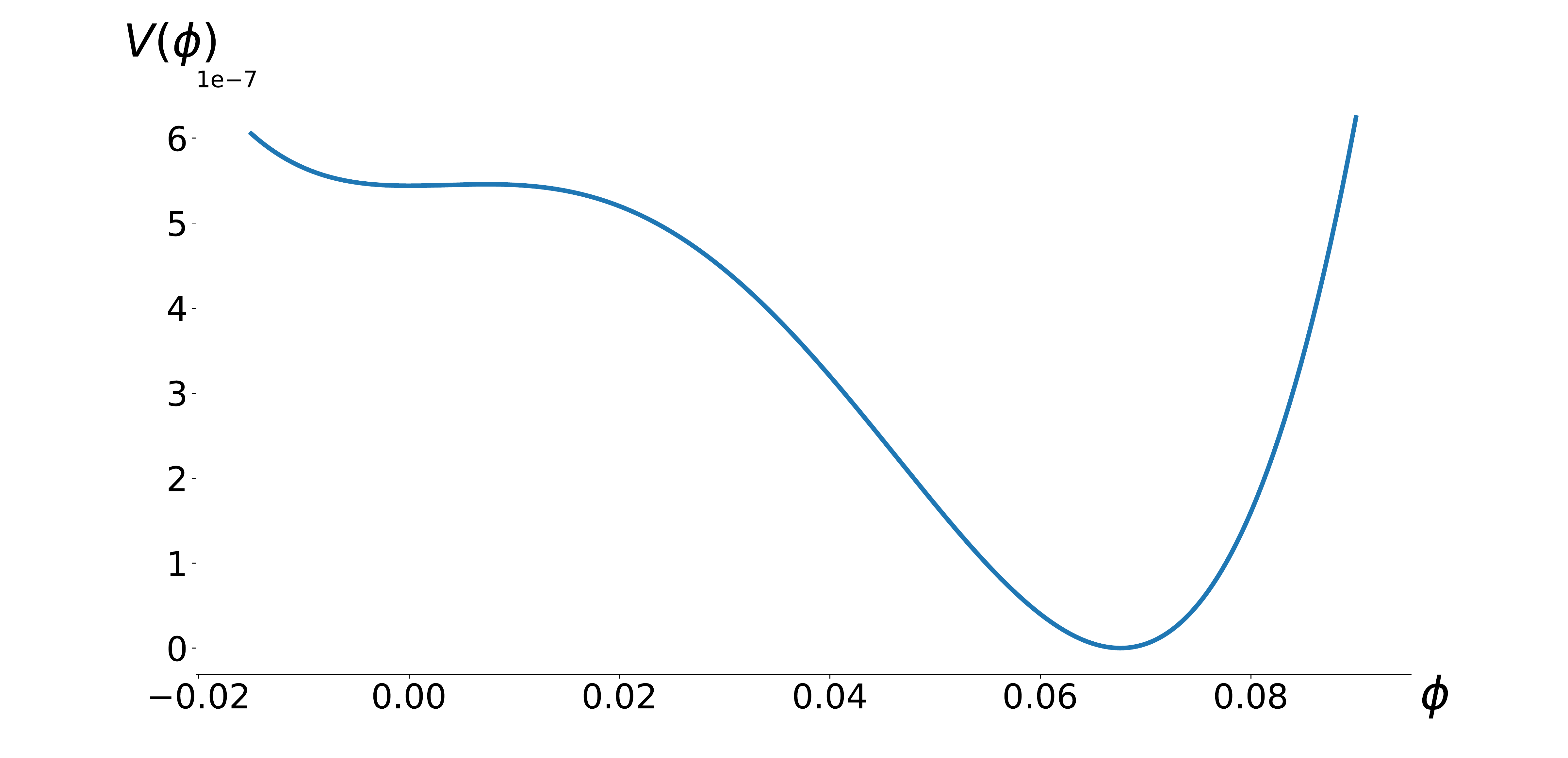}
\caption{\label{up-po-2}{Potential with $V_d \sim V_b \gg V_p$}}
\end{figure}

As discussed in the previous subsection, it would be interesting to numerically investigate 
the case where the expansion rate of the daughter vacuum is much larger than that in the 
parent vacuum. The potential we consider for this case is given by,
\beq
 V(\phi) = 10^{-8} [(100\phi) ^2 - (100\phi)^3 + 0.1(100\phi)^4] + 5.44\times 10^{-7},
\eeq 
which is shown in Fig. \ref{up-po-2}. In this potential $m^{-1} \approx 25$, 
$H_p^{-1} \approx 3.7 \times 10^{4}$ and $H_d^{-1} \approx 470$.  As before, the field is initially in the true 
vacuum, $\phi(t_0,r)\approx 0.068$ and the fluctuation parameters are $l= 2.7H_d^{-1} \ll H_p^{-1}$ 
and $\dot{\phi}_0 = -1.3\sqrt{2V_b}$.

\begin{figure}
   \centering
\includegraphics[scale=0.5]{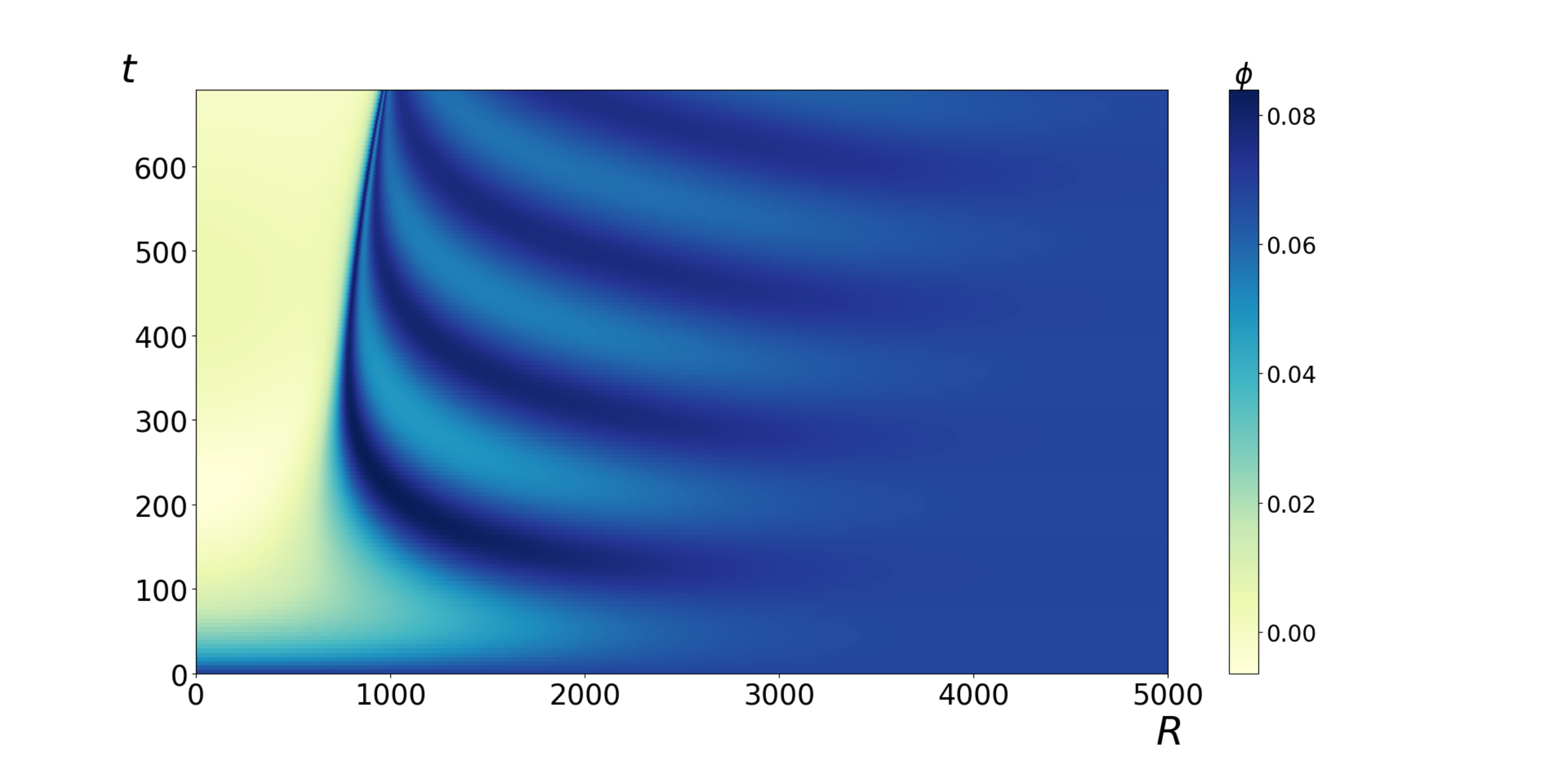}
\caption{\label{up-2}{Evolution of the scalar field under the potential Fig. \ref{up-po-2}, with 
$l= 2.7H_d^{-1}$ and $\dot{\phi}_0 = -1.3\sqrt{2V_b}$. The horizontal axis is the area radius $R$ 
and the vertical axis is time $t$. {A false vacuum bubble forms and expands.}}}
\end{figure}

We show in Fig. \ref{up-2} the field evolution in this case where the fluctuation 
is strong enough to bring the field over the barrier and create a bubble of the false vacuum.
We note that the size of this bubble is of the order of $H_d^{-1}$ and much smaller than 
$H_p^{-1}$ and once formed it does not collapse. This example shows that upward transitions
of this type indeed seem to happen in a curved spacetime.

\begin{figure}
   \centering
\includegraphics[scale=0.7]{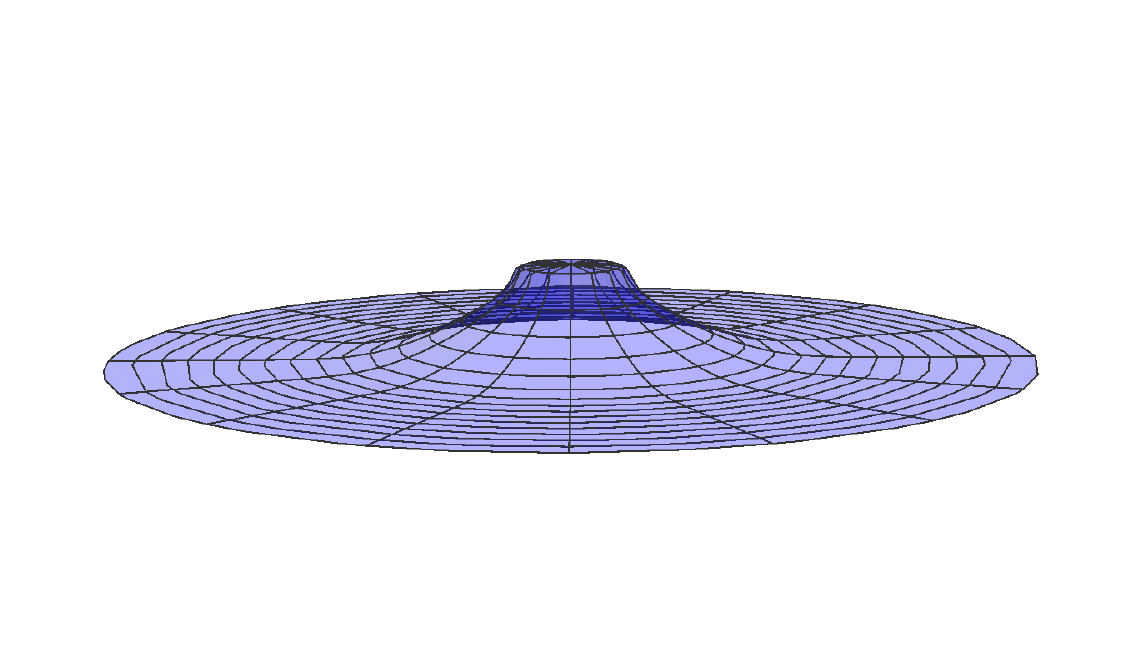}\\
\includegraphics[scale=0.7]{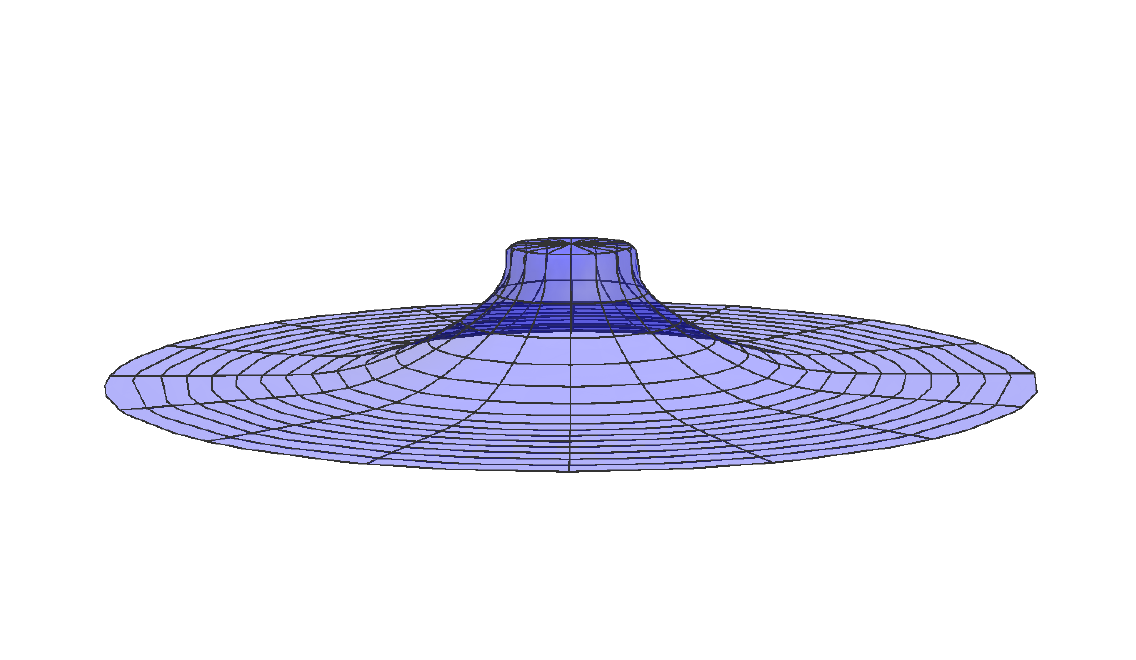}\\
\includegraphics[scale=0.7]{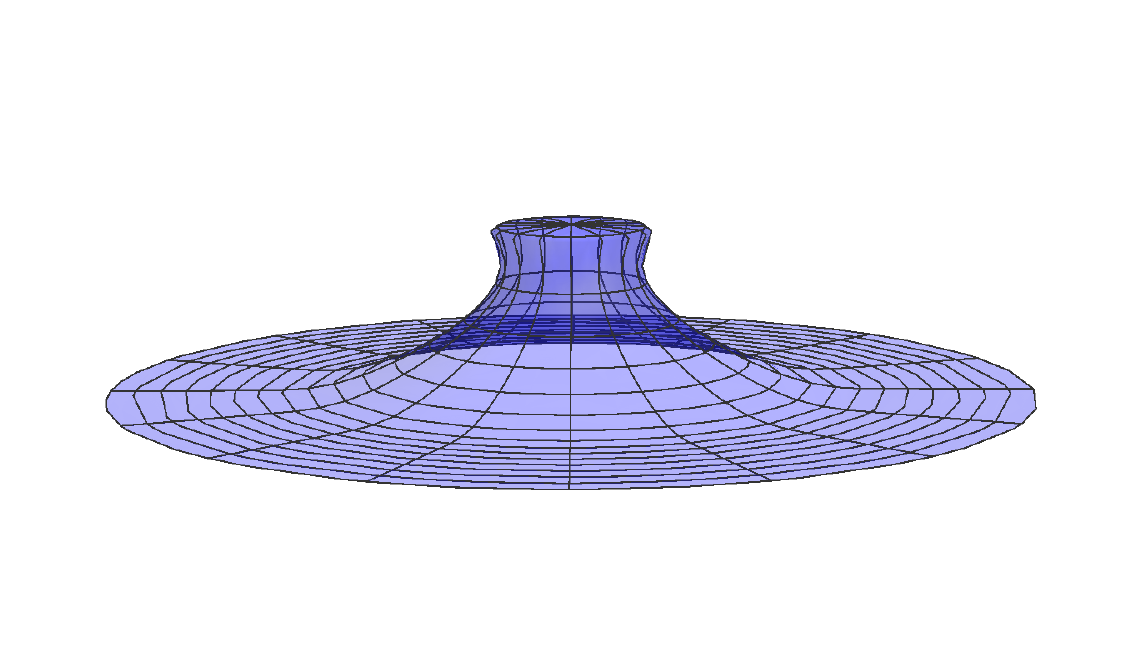}\\
\includegraphics[scale=0.7]{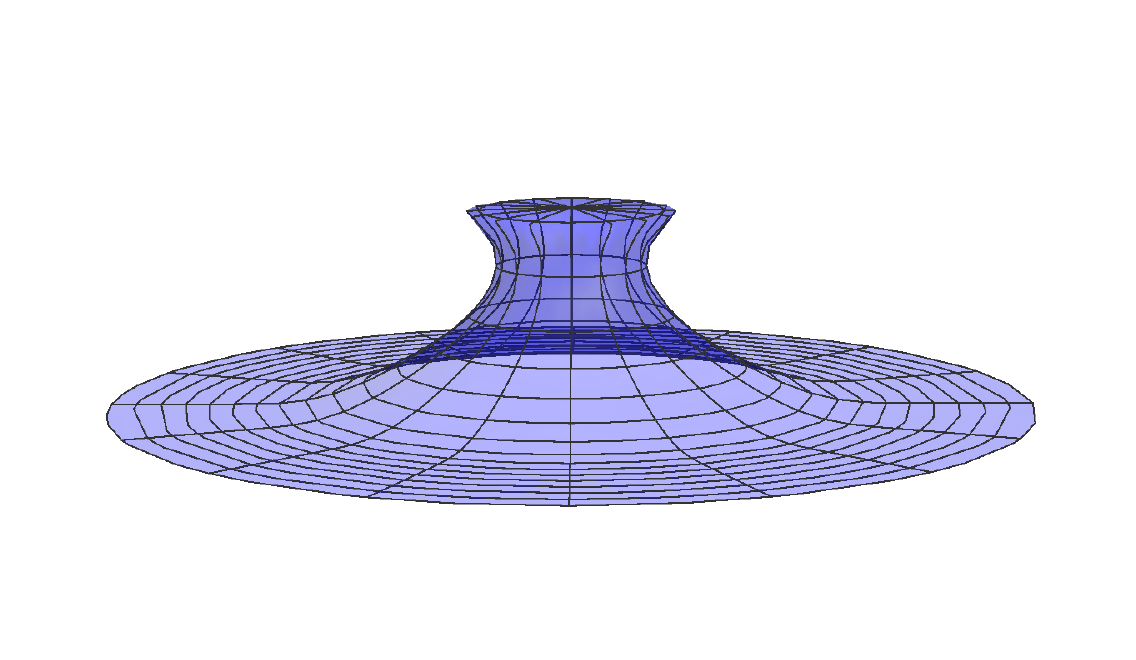}
\protect\caption{\label{wormhole}Embedding diagrams at different moments showing how the 
wormhole develops. The flat-looking region at the top and
the bottom area represent the daughter and parent regions respectively.
The ring that encircles the cap is the bubble wall. The
arc length from the cap center along longitude is the proper
radius $d(r)\equiv\int_{0}^{r}B(\tilde{r})d\tilde{r}$, while
the radius of a latitude circle is the area radius $R$.}
\end{figure}

The exponentially growing stage is not shown in Fig. \ref{up-2} due to the formation of a wormhole geometry. 
Since the expansion rate of the false vacuum is much larger than that in the background, the bubble 
inflates into a baby universe, which is connected to the parent universe by a wormhole. After the 
wormhole is formed, the area radius $R$ is no longer a monotonic function of the coordinate $r$ 
(hence it can no longer serve as the horizontal axis in Fig. \ref{up-2}) but has a local minimum, 
which is the wormhole throat. This throat would eventually close off and the wormhole would turn 
into a Schwarzschild-de Sitter black hole for an observer in the parent universe. In the above example, our simulation shows that the resulting black hole has mass $M \approx 750$. This is consistent with the estimate (\ref{MBH}).\footnote{We have checked examples with a considerable range of $H_d^{-1}$, all of which agree with Eq. (\ref{MBH}) by order of magnitude.}  In Fig. \ref{wormhole} 
we show the wormhole formation by its embedding diagrams at four different moments.

\section{Conclusions}

We used analytic and numerical methods to investigate flyover vacuum transitions in flat and 
dS space.  Our numerical simulations indicate that the dynamics of bubble nucleation and early 
evolution in these transitions can be rather different from the standard picture.  The difference 
is most pronounced for thin-wall bubbles, where the transition region undergoes a number 
of oscillations between true and false vacuum, until a true vacuum shell is formed which expands 
both inwards and outwards.  This suggests that flyover transitions provide another channel 
of vacuum decay, different from quantum tunneling.  Our analytic estimates of flyover decay 
rates in flat space and of downward decays in dS space have the same parametric dependence 
as given by the instanton method for tunneling transitions, with a numerical coefficient in the
 exponent different by a factor ${\cal O}(1-10)$.  The estimates suggest that the flyover rate 
is smaller than the tunneling rate, although their accuracy is not sufficient to establish this 
with certainty.  If the flyover decays are indeed subdominant, the standard results for the 
vacuum decay rate would still apply.

On the other hand, for upward transitions in dS space the flyover decay rate is parametrically 
different from tunneling and can be higher by many orders of magnitude.  This is an interesting 
result, which implies in particular that the detailed balance condition (\ref{kappaupdown}) is 
violated.  If correct, this violation would be a significant deviation from the conventional 
picture of quantum dS space as a thermal state \cite{Dyson:2002pf}.

Another important difference from the standard picture is that unlike Coleman's or the CdL 
bubbles, flyover vacuum transitions are not symmetric with respect to Lorentz or de Sitter 
boosts.  There is a preferred frame where the fluctuation is spherically symmetric, which 
can be regarded as the rest frame of bubble nucleation.
This is not necessarily a problem, since a metastable vacuum always has a preferred frame, 
which is set by the way the vacuum was created and which manifests itself for example by 
the "persistence of memory" effect \cite{Garriga:2006hw}.

In most of our analysis we assumed that the tunneling barrier height is small compared to 
the parent vacuum energy density, $V_b\ll V_p$, to ensure that gravitational back-reaction 
of the initial fluctuation is small.  We also considered upward dS transitions where the daughter 
vacuum has a much higher energy density than the parent vacuum, $V_d\gg V_p$.  Then we 
must also have $V_b\gg V_p$ and the transition rate cannot be reliably estimated.  We 
expect, however, the rate to be nonzero and we performed numerical simulations of flyover 
transitions in this case.  We find that after the transition the rapidly inflating daughter region 
is connected to the parent vacuum by a wormhole.  The wormhole closes off in about a light 
crossing time, resulting in an inflating daughter bubble contained in a baby universe inside of a black hole.

In this paper we focused on flyover transitions for which an alternative CdL tunneling transition 
is also possible.  We note however that there are generic situations where CdL instantons do 
not exist.  For example, tunneling between two dS vacua may not be possible if there is an 
Anti de Sitter (AdS) or Minkowski vacuum between them.  This could actually be the case for most dS vacuum 
pairs in the Landscape \cite{BDV}.  In such cases a flyover transition may be the only available 
vacuum decay channel. These types of transitions could also be relevant in the calculation
of the spectrum of primordial black holes produced during inflation in 
scenarios previously discussed in Refs.~\cite{Garriga:2015fdk,Deng:2017uwc}.

We finally emphasize that our approach to estimating the transition rates is heuristic in 
nature.  We used quantum field theory to estimate the probability of the initial fluctuation 
and then evolved the fluctuation using the classical field equations.  Justification of this 
method from first principles and determining the limits of its applicability remain problems 
for future research.  Some steps in this direction, using the Wigner distribution functions, 
were taken in Ref.~\cite{Braden:2018tky,Hertzberg:2019wgx}.

\section{Acknowledgements}

We are grateful to Adam Brown, Larry Ford, Haiyun Huang, Matthew Johnson, Matthew Kleban and Ken Olum for very useful discussions.  J. J. B.-P. is 
supported in part by the Spanish Ministry MINECO grant (FPA2015-64041-C2-1P), the MCIU/AEI/FEDER
grant (PGC2018-094626-B-C21), the Basque Government grant (IT-979-16) and the Basque 
Foundation for Science (IKERBASQUE). H. D. and A. V. are supported by the National Science 
Foundation under grant PHY-1820872.

\bibliography{flyover}

\begin{thebibliography}{22}
\expandafter\ifx\csname natexlab\endcsname\relax\def\natexlab#1{#1}\fi
\expandafter\ifx\csname bibnamefont\endcsname\relax
  \def\bibnamefont#1{#1}\fi
\expandafter\ifx\csname bibfnamefont\endcsname\relax
  \def\bibfnamefont#1{#1}\fi
\expandafter\ifx\csname citenamefont\endcsname\relax
  \def\citenamefont#1{#1}\fi
\expandafter\ifx\csname url\endcsname\relax
  \def\url#1{\texttt{#1}}\fi
\expandafter\ifx\csname urlprefix\endcsname\relax\def\urlprefix{URL }\fi
\providecommand{\bibinfo}[2]{#2}
\providecommand{\eprint}[2][]{\url{#2}}

\bibitem[{\citenamefont{Ellis et~al.}(1990)\citenamefont{Ellis, Linde, and
  Sher}}]{Ellis:1990bv}
\bibinfo{author}{\bibfnamefont{J.~R.} \bibnamefont{Ellis}},
  \bibinfo{author}{\bibfnamefont{A.~D.} \bibnamefont{Linde}}, \bibnamefont{and}
  \bibinfo{author}{\bibfnamefont{M.}~\bibnamefont{Sher}},
  \bibinfo{journal}{Phys. Lett.} \textbf{\bibinfo{volume}{B252}},
  \bibinfo{pages}{203} (\bibinfo{year}{1990}).

\bibitem[{\citenamefont{Linde}(1992)}]{Linde:1991sk}
\bibinfo{author}{\bibfnamefont{A.~D.} \bibnamefont{Linde}},
  \bibinfo{journal}{Nucl. Phys.} \textbf{\bibinfo{volume}{B372}},
  \bibinfo{pages}{421} (\bibinfo{year}{1992}), \eprint{hep-th/9110037}.

\bibitem[{\citenamefont{Brown and Dahlen}(2011)}]{Brown:2011ry}
\bibinfo{author}{\bibfnamefont{A.~R.} \bibnamefont{Brown}} \bibnamefont{and}
  \bibinfo{author}{\bibfnamefont{A.}~\bibnamefont{Dahlen}},
  \bibinfo{journal}{Phys. Rev. Lett.} \textbf{\bibinfo{volume}{107}},
  \bibinfo{pages}{171301} (\bibinfo{year}{2011}), \eprint{1108.0119}.

\bibitem[{\citenamefont{Braden et~al.}(2018)\citenamefont{Braden, Johnson,
  Peiris, Pontzen, and Weinfurtner}}]{Braden:2018tky}
\bibinfo{author}{\bibfnamefont{J.}~\bibnamefont{Braden}},
  \bibinfo{author}{\bibfnamefont{M.~C.} \bibnamefont{Johnson}},
  \bibinfo{author}{\bibfnamefont{H.~V.} \bibnamefont{Peiris}},
  \bibinfo{author}{\bibfnamefont{A.}~\bibnamefont{Pontzen}}, \bibnamefont{and}
  \bibinfo{author}{\bibfnamefont{S.}~\bibnamefont{Weinfurtner}}
  (\bibinfo{year}{2018}), \eprint{1806.06069}.

\bibitem[{\citenamefont{Huang and Ford}(2019)}]{Huang}
\bibinfo{author}{\bibfnamefont{H.}~\bibnamefont{Huang}} \bibnamefont{and}
  \bibinfo{author}{\bibfnamefont{L.~H.} \bibnamefont{Ford}},
  \bibinfo{journal}{work in progress}  (\bibinfo{year}{2019}),
  \bibinfo{note}{seminar given by H. Huang at Tufts University, April 2018.}

\bibitem[{\citenamefont{Hertzberg and Yamada}(2019)}]{Hertzberg:2019wgx}
\bibinfo{author}{\bibfnamefont{M.~P.} \bibnamefont{Hertzberg}}
  \bibnamefont{and} \bibinfo{author}{\bibfnamefont{M.}~\bibnamefont{Yamada}}
  (\bibinfo{year}{2019}), \eprint{1904.08565}.

\bibitem[{\citenamefont{Coleman}(1977)}]{Coleman:1977py}
\bibinfo{author}{\bibfnamefont{S.~R.} \bibnamefont{Coleman}},
  \bibinfo{journal}{Phys. Rev.} \textbf{\bibinfo{volume}{D15}},
  \bibinfo{pages}{2929} (\bibinfo{year}{1977}), \bibinfo{note}{[Erratum: Phys.
  Rev.D16,1248(1977)]}.

\bibitem[{\citenamefont{Coleman and De~Luccia}(1980)}]{Coleman:1980aw}
\bibinfo{author}{\bibfnamefont{S.~R.} \bibnamefont{Coleman}} \bibnamefont{and}
  \bibinfo{author}{\bibfnamefont{F.}~\bibnamefont{De~Luccia}},
  \bibinfo{journal}{Phys. Rev.} \textbf{\bibinfo{volume}{D21}},
  \bibinfo{pages}{3305} (\bibinfo{year}{1980}).

\bibitem[{\citenamefont{Lee and Weinberg}(1987)}]{Lee:1987qc}
\bibinfo{author}{\bibfnamefont{K.-M.} \bibnamefont{Lee}} \bibnamefont{and}
  \bibinfo{author}{\bibfnamefont{E.~J.} \bibnamefont{Weinberg}},
  \bibinfo{journal}{Phys. Rev.} \textbf{\bibinfo{volume}{D36}},
  \bibinfo{pages}{1088} (\bibinfo{year}{1987}).

\bibitem[{\citenamefont{Hawking and Moss}(1982)}]{Hawking:1981fz}
\bibinfo{author}{\bibfnamefont{S.~W.} \bibnamefont{Hawking}} \bibnamefont{and}
  \bibinfo{author}{\bibfnamefont{I.~G.} \bibnamefont{Moss}},
  \bibinfo{journal}{Phys. Lett.} \textbf{\bibinfo{volume}{110B}},
  \bibinfo{pages}{35} (\bibinfo{year}{1982}), \bibinfo{note}{[Adv. Ser.
  Astrophys. Cosmol.3,154(1987)]}.

\bibitem[{\citenamefont{Dyson et~al.}(2002)\citenamefont{Dyson, Kleban, and
  Susskind}}]{Dyson:2002pf}
\bibinfo{author}{\bibfnamefont{L.}~\bibnamefont{Dyson}},
  \bibinfo{author}{\bibfnamefont{M.}~\bibnamefont{Kleban}}, \bibnamefont{and}
  \bibinfo{author}{\bibfnamefont{L.}~\bibnamefont{Susskind}},
  \bibinfo{journal}{JHEP} \textbf{\bibinfo{volume}{10}}, \bibinfo{pages}{011}
  (\bibinfo{year}{2002}), \eprint{hep-th/0208013}.

\bibitem[{\citenamefont{Sarid}(1998)}]{Sarid:1998sn}
\bibinfo{author}{\bibfnamefont{U.}~\bibnamefont{Sarid}},
  \bibinfo{journal}{Phys. Rev.} \textbf{\bibinfo{volume}{D58}},
  \bibinfo{pages}{085017} (\bibinfo{year}{1998}), \eprint{hep-ph/9804308}.

\bibitem[{\citenamefont{Dine and Paban}(2015)}]{Dine:2015ioa}
\bibinfo{author}{\bibfnamefont{M.}~\bibnamefont{Dine}} \bibnamefont{and}
  \bibinfo{author}{\bibfnamefont{S.}~\bibnamefont{Paban}},
  \bibinfo{journal}{JHEP} \textbf{\bibinfo{volume}{10}}, \bibinfo{pages}{088}
  (\bibinfo{year}{2015}), \eprint{1506.06428}.

\bibitem[{\citenamefont{Linde}(1983)}]{Linde:1981zj}
\bibinfo{author}{\bibfnamefont{A.~D.} \bibnamefont{Linde}},
  \bibinfo{journal}{Nucl. Phys.} \textbf{\bibinfo{volume}{B216}},
  \bibinfo{pages}{421} (\bibinfo{year}{1983}), \bibinfo{note}{[Erratum: Nucl.
  Phys.B223,544(1983)]}.

\bibitem[{\citenamefont{Basu et~al.}(1991)\citenamefont{Basu, Guth, and
  Vilenkin}}]{Basu:1991ig}
\bibinfo{author}{\bibfnamefont{R.}~\bibnamefont{Basu}},
  \bibinfo{author}{\bibfnamefont{A.~H.} \bibnamefont{Guth}}, \bibnamefont{and}
  \bibinfo{author}{\bibfnamefont{A.}~\bibnamefont{Vilenkin}},
  \bibinfo{journal}{Phys. Rev.} \textbf{\bibinfo{volume}{D44}},
  \bibinfo{pages}{340} (\bibinfo{year}{1991}).

\bibitem[{\citenamefont{Garriga et~al.}(2016)\citenamefont{Garriga, Vilenkin,
  and Zhang}}]{Garriga:2015fdk}
\bibinfo{author}{\bibfnamefont{J.}~\bibnamefont{Garriga}},
  \bibinfo{author}{\bibfnamefont{A.}~\bibnamefont{Vilenkin}}, \bibnamefont{and}
  \bibinfo{author}{\bibfnamefont{J.}~\bibnamefont{Zhang}},
  \bibinfo{journal}{JCAP} \textbf{\bibinfo{volume}{1602}}, \bibinfo{pages}{064}
  (\bibinfo{year}{2016}), \eprint{1512.01819}.

\bibitem[{\citenamefont{Deng and Vilenkin}(2017)}]{Deng:2017uwc}
\bibinfo{author}{\bibfnamefont{H.}~\bibnamefont{Deng}} \bibnamefont{and}
  \bibinfo{author}{\bibfnamefont{A.}~\bibnamefont{Vilenkin}},
  \bibinfo{journal}{JCAP} \textbf{\bibinfo{volume}{1712}}, \bibinfo{pages}{044}
  (\bibinfo{year}{2017}), \eprint{1710.02865}.

\bibitem[{\citenamefont{Farhi et~al.}(1990)\citenamefont{Farhi, Guth, and
  Guven}}]{Farhi:1989yr}
\bibinfo{author}{\bibfnamefont{E.}~\bibnamefont{Farhi}},
  \bibinfo{author}{\bibfnamefont{A.~H.} \bibnamefont{Guth}}, \bibnamefont{and}
  \bibinfo{author}{\bibfnamefont{J.}~\bibnamefont{Guven}},
  \bibinfo{journal}{Nucl. Phys.} \textbf{\bibinfo{volume}{B339}},
  \bibinfo{pages}{417} (\bibinfo{year}{1990}).

\bibitem[{\citenamefont{Garriga and Megevand}(2004)}]{Garriga:2004nm}
\bibinfo{author}{\bibfnamefont{J.}~\bibnamefont{Garriga}} \bibnamefont{and}
  \bibinfo{author}{\bibfnamefont{A.}~\bibnamefont{Megevand}},
  \bibinfo{journal}{Int. J. Theor. Phys.} \textbf{\bibinfo{volume}{43}},
  \bibinfo{pages}{883} (\bibinfo{year}{2004}), \eprint{hep-th/0404097}.

\bibitem[{\citenamefont{Aguirre and Johnson}(2006)}]{Aguirre:2005nt}
\bibinfo{author}{\bibfnamefont{A.}~\bibnamefont{Aguirre}} \bibnamefont{and}
  \bibinfo{author}{\bibfnamefont{M.~C.} \bibnamefont{Johnson}},
  \bibinfo{journal}{Phys. Rev.} \textbf{\bibinfo{volume}{D73}},
  \bibinfo{pages}{123529} (\bibinfo{year}{2006}), \eprint{gr-qc/0512034}.

\bibitem[{\citenamefont{Garriga et~al.}(2007)\citenamefont{Garriga, Guth, and
  Vilenkin}}]{Garriga:2006hw}
\bibinfo{author}{\bibfnamefont{J.}~\bibnamefont{Garriga}},
  \bibinfo{author}{\bibfnamefont{A.~H.} \bibnamefont{Guth}}, \bibnamefont{and}
  \bibinfo{author}{\bibfnamefont{A.}~\bibnamefont{Vilenkin}},
  \bibinfo{journal}{Phys. Rev.} \textbf{\bibinfo{volume}{D76}},
  \bibinfo{pages}{123512} (\bibinfo{year}{2007}), \eprint{hep-th/0612242}.

\bibitem[{\citenamefont{Blanco-Pillado
  et~al.}(2019)\citenamefont{Blanco-Pillado, Deng, and Vilenkin}}]{BDV}
\bibinfo{author}{\bibfnamefont{J.~J.} \bibnamefont{Blanco-Pillado}},
  \bibinfo{author}{\bibfnamefont{H.}~\bibnamefont{Deng}}, \bibnamefont{and}
  \bibinfo{author}{\bibfnamefont{A.}~\bibnamefont{Vilenkin}},
  \bibinfo{journal}{work in progress}  (\bibinfo{year}{2019}).

\end{thebibliography}

\appendix
\section{}

Here we find the rms magnitude of field velocity fluctuations in Minkowski and de Sitter space.  

\subsection{Minkowski space}

A free scalar field of mass $m$ in Minkowski space is represented by a field operator
\beq
{\hat\phi}({\bf x},t)=(2\pi)^{-3/2} \int d^3 k \left[{\hat a}_{\bf k}\psi_k(t)e^{i{\bf kx}} +
 {\hat a}^\dagger_{\bf k}\psi^*_k(t)e^{-i{\bf kx}}\right],
\eeq
where the annihilation and creation operators ${\hat a}, {\hat a}^\dagger$ satisfy the standard 
commutation relations, 
$\psi_k(t) = (2E_k)^{-1/2} e^{-iE_k t}$, $E_k=\sqrt{k^2+m^2}$ and $k=|{\bf k}|$.
We define a velocity operator smeared over distance scale $l$: 
\beq
{\dot{\hat\phi}}_l({\bf x},t)= (2\pi d^2)^{-3/2} \int d^3 x' {\dot{\hat\phi}}({\bf x},t) \exp \left(-\frac{|{\bf x}-{\bf x}'|^2} {2l^2}\right).
\label{smeared}
\eeq
We are interested in the quantity
\beq
\langle {\dot\phi}_l^2\rangle\equiv \langle 0|{\dot{\hat\phi}}_l^2({\bf x},t)|0\rangle = \frac{1}{4\pi^2} \int_0^\infty e^{-k^2 l^2} E_k k^2 dk = 
\frac{(ml)^2 e^{(ml)^2}}{16 \pi^2 l^4} K_1 ( ml/2)
\label{variance}
\eeq
where we have introduced $K_1(x)$ which denotes a modified Bessel function. For $ml\gg 1$ we can 
approximate $E_k\approx m$; this gives in this limit,
\beq
\langle{\dot{\phi}}_l^2 \rangle \approx \frac{m}{16\pi^{3/2} l^3}.
\label{vaccontr}
\eeq
In the opposite limit, $ml\ll 1$, $E_k\approx k$ and we have
\beq
\langle{\dot{\phi}}_l^2 \rangle \approx\frac{1}{8\pi^2 l^4}.
\eeq

\subsection{Finite temperature}

In a thermal state at temerature $T$ we have $\langle a^\dagger a\rangle=n_k$ and 
Eq.~(\ref{variance}) is replaced by
\beq
\langle {\dot\phi}_l^2\rangle= \frac{1}{4\pi^2} \int_0^\infty e^{-k^2 l^2} (1+2n_k) E_k k^2 dk = \langle {\dot\phi}_l^2\rangle_v +\langle {\dot\phi}_l^2\rangle_T.
\eeq
Here, angular brackets indicate averaging with a thermal density matrix, 
$n_k=\left(e^{E_k /T}-1\right)^{-1}$ is the Bose distribution function, and the two terms on 
the right-hand side represent vacuum and thermal contributions.  The vacuum contribution is 
the same as at $T=0$ and the thermal contribution is
\beq
\langle {\dot\phi}_l^2\rangle_T= \frac{1}{2\pi^2} \int_0^\infty e^{-k^2 l^2}\frac{ E_k k^2 dk}{e^{E_k /T}-1}. 
\eeq

For $T\ll m$ the thermal contribution is exponentially suppressed, $\langle {\dot\phi}_l^2\rangle_T \propto e^{-m/T}$.  
On the other hand, at $T\gg m, 1/l$ we have
\beq
\langle {\dot\phi}_l^2\rangle_T\approx \frac{T}{2\pi^2} \int_0^\infty e^{-k^2 l^2} k^2 dk  = \frac{T}{8\pi^{3/2} l^3}. 
\eeq
This is much larger than the vacuum contribution (\ref{vaccontr}); hence thermal fluctuations dominate in this regime.

\subsection{de Sitter space}

We now consider a scalar field in dS space,
\beq
ds^2 = a^2(\eta) \left(d\eta^2 - d{\bf x}^2\right),
\eeq 
where $a(\eta)=-(H\eta)^{-1}$.  The time variable $\eta$ varies in the range $-\infty<\eta< 0$ and 
is related to the proper time $t$ by $dt = -a(\eta)d\eta$.  A free scalar field operator in dS space can 
be represented as
\beq
{\hat\phi}(x)=(2\pi)^{-3/2} \int d^3 k \left[{\hat a}_{\bf k}\psi_k(\eta)e^{i{\bf kx}} + {\hat a}^\dagger_{\bf k}\psi^*_k(\eta)e^{-i{\bf kx}}\right],
\eeq
Assuming that the field is in a Bunch-Davies vacuum state, the mode functions are given by
\beq
\psi_k(\eta)=\frac{\sqrt{\pi} H}{2} e^{\frac{i\pi\mu}{2}} (-\eta)^{3/2} H_\mu^{(1)} (-k\eta) ,
 \eeq
 where $H_\mu^{(1)}(x)$ are the Hankel functions and
 \beq
 \mu =\sqrt{\frac{9}{4}-\frac{m^2}{H^2}}.
 \eeq
 
The smeared velocity operator is given by 
\beq
{\dot{\hat\phi}}_l({\bf x},\eta)= (2\pi d^2)^{-3/2} \int d^3 x' {\dot{\hat\phi}}({\bf x},\eta) \exp \left(-\frac{|{\bf x}-{\bf x}'|^2} {2H^2\eta^2 l^2}\right) ,
\eeq
where dots stand for derivatives with respect to the proper time $t$ and $l$ is the physical smearing length.
The rms field velocity fluctuation is characterized by the vacuum expectation value
\beq
{\dot\phi}^2_l \equiv \langle 0|{\dot{\hat \phi}}_l^2({\bf x},\eta)|0\rangle = \frac{H^2\eta^2}{2\pi^2} \int_0^\infty dk k^2 \left| \frac{d\psi_k}{d\eta}\right|^2 e^{-{H^2 \eta^2 k^2 l^2}}.
\label{dotphil}
\eeq

We are interested in the case when $m\gg H$.  Then $\mu \approx im/H \equiv i\nu$.  The mode 
functions in this regime can be found using the integral representation of the Hankel functions,
\beq
H_{i\nu}^{(1)}(x) = \frac{1}{i\pi} e^{\frac{\pi\nu}{2}} \int_{-\infty}^\infty d\xi e^{i(x\cosh \xi-\nu\xi)}.
\label{Hankel}
\eeq
For $\nu\gg 1$ the integral in (\ref{Hankel}) can be evaluated using the stationary phase approximation; this gives
\beq
H_{i\nu}^{(1)}(x) \approx \sqrt{\frac{2}{i\pi}} e^{\frac{\pi\nu}{2}} (x^2+\nu^2)^{-1/4} e^{if(x)} ,
\label{Hankelx}
\eeq
where
\beq
f(x)= \sqrt{x^2+\nu^2} -\nu \ln\frac{x+\sqrt{x^2+\nu^2}} {\nu}.
\label{fx}
\eeq

The integral over $k$ in Eq.~(\ref{dotphil}) is dominated by the values $k \sim 1/H\eta l$.  For $ml\ll 1$ 
this corresponds to the argument of the Hankel function $k|\eta| \gg \nu$.  Then
\beq
\psi_k(\eta)\approx \frac{H |\eta|}{\sqrt{2ik}} e^{-ik\eta}
\eeq
and
\beq
\langle{\dot\phi}^2_l \rangle\approx \frac{H^4\eta^4}{4\pi^2} \int_0^\infty dk k^3 e^{-{H^2 \eta^2 k^2 l^2}} = \frac{1}{8\pi^2 l^4}.
\eeq
As one might expect, ${\dot\phi}^2_l$ is independent of time $\eta$.  In the opposite regime, $ml\gg 1$, we have
\beq
\psi_k(\eta)\approx \frac{H}{\sqrt{2i\nu}} (-\eta)^{3/2} e^{i\nu\left(1+\ln\frac{k|\eta|}{2\nu}\right)} 
\label{approxmode}
\eeq
and
\beq
\langle{\dot\phi}^2_l\rangle \approx \frac{m}{16\pi^{3/2} l^3}.
\eeq
Somewhat surprisingly, these results are the same as in Minkowski space in both limiting cases.

As an independent check of Eq.~(\ref{approxmode}) we note that it can be rewritten as
\beq
\psi_k(t) \approx \frac{1}{\sqrt{2m}} [a(t)]^{-3/2} e^{-imt +i\alpha_k},
\eeq
where $a(t)=e^{Ht}=(H|\eta)^{-1}$ is the scale factor and $\alpha_k$ is a constant phase.  This form 
of the mode functions can also be obtained by solving the field equation
\beq
{\ddot\psi}_k +3H{\dot\psi}_k +k^2 a^{-2}(t)\psi_k +m^2 \psi_k = 0.
\label{modeeq}
\eeq
For $ml\gg 1$ the relevant mode functions are the ones with $k/a \ll m$, so the gradient term 
in (\ref{modeeq}) can be neglected and the solution is (for $m\gg H$)
\beq
\psi_k(t)\approx A_k a^{-3/2} e^{-imt}.
\eeq
The normalization constant $A_k$ can be found from the condition
\beq
\psi_k {\dot\psi}^*_k -\psi^*_k {\dot\psi}_k = ia^{-3}.
\eeq
This gives $|A_k|\approx (2m)^{-1/2}$, and thus we recover Eq.~(\ref{approxmode}).

We have also verified numerically that Eqs.~(\ref{Hankelx}), (\ref{fx}) give a good fit to Hankel 
functions of large imaginary order.

\section{}

In this appendix we outline the setup of our simulations used to illustrate the upward transitions between 
dS vacua when gravity is taken into account. 

Consider a general spherically symmetric metric
\beq
ds^2=-dt^2+B^2 dr^2+R^2 d\Omega^2,
\eeq
where $d\Omega^2$ is the metric on a unit sphere and $B$ and $R$ are functions of the coordinates 
$t$ and $r$. For later convenience, we define
\beq
U = \dot{R},~~~~~  \Gamma = \frac{R^\prime}{B},~~~~~ K=\frac{\dot{B}}{B}+\frac{2\dot{R}}{R}.
\eeq
The Lagrangian density of a scalar field $\phi$ with potential $V(\phi)$ is
\beq
\mathcal{L}=\frac{1}{2}(\partial \phi)^2 - V(\phi),
\eeq
and the corresponding energy-momentum tensor is given by
\beq
-T^{0}_{\ 0} = \frac{1}{2}\dot{\phi}^2 + \frac{1}{2} \left(\frac{\phi{\prime}}{B}\right)^2 + V(\phi),
\eeq
\beq
T^{1}_{\ 1} = \frac{1}{2}\dot{\phi}^2 + \frac{1}{2} \left(\frac{\phi{\prime}}{B}\right)^2 - V(\phi),
\eeq
\beq
T^{2}_{\ 2} = \frac{1}{2}\dot{\phi}^2 - \frac{1}{2} \left(\frac{\phi{\prime}}{B}\right)^2 - V(\phi),
\eeq
\beq
T^{0}_{\ 1} = \dot{\phi}\phi^{\prime}.
\eeq

Using Einstein's equations and the scalar field equation we obtain,
\begin{equation}
\dot{K}=-\left(K-\frac{2U}{R}\right)^{2}-2\left(\frac{U}{R}\right)^{2}-4\pi(T_{\ 1}^{1}+2T_{\ 2}^{2}-T_{\ 0}^{0})\label{K}
\end{equation}
\begin{equation}
\dot{U}=-\frac{1-\Gamma^{2}+U^{2}}{2R}-4\pi RT_{\ 1}^{1}\label{U}
\end{equation}
\begin{equation}
\dot{\Gamma}=-\frac{4\pi RT_{\ 1}^{0}}{B}\label{G}
\end{equation}
\begin{equation}
\dot{B}=B\left(K-\frac{2U}{R}\right)\label{dK}
\end{equation}
\begin{equation}
\dot{R}=U\label{dU}
\end{equation}
\begin{equation}
\ddot{\phi}=-K\dot{\phi}+\frac{1}{BR^{2}}\left(\frac{R^{2}}{B}\phi^{\prime}\right)^{\prime}-\partial_{\phi}V\label{phi}
\end{equation}
which are the evolution equations for the functions
$K,\ U,\ \Gamma,\ B,\ R$ and $\phi$.

To evolve this system we need initial and boundary conditions of the equations. The initial profile of the 
scalar field and that of the field velocity are respectively set to be
\beq
\phi(r)=\phi_p,~~~~~ \dot{\phi}(r)=\dot{\phi}_0 e^{-\frac{r^2}{2l^2}},
\eeq
where $\phi_p$ is the field value at true (parent) vaccum and $l$ characterizes the width of the velocity 
fluctuation. The initial energy density is given by $\rho(r)\equiv -T_{\ 0}^{0}=\dot{\phi}^2_0 e^{-\frac{r^2}{l^2}}/2+V(\phi_p)$. 

As for the metric functions at the initial time, we set $B(r)=1$ and $R(r)=r$. Then by definition we have 
$\Gamma(r)=1$. The mass enclosed by a certain sphere is characterized by the Misner-Sharp mass, 
which in our coordinates is
\beq
M(r)=\frac{R}{2}\left(1-\Gamma^{2}+U^{2}\right).
\eeq
From $G_{00}\propto T_{00}$, we have $M^{\prime}(r)=4\pi\rho R^{2}R^{\prime}$, which gives
\beq
M(r)=4\pi\int\rho r^{2}dr.
\eeq
This can be found by numerical integration. Then by the definition of the Misner-Sharp mass, the initial 
profile of $U$ is
\beq
U(r)=\sqrt{\frac{2M}{r}},
\eeq
where we have used $\Gamma = 1$. From $G_{01}\propto T_{01}$, it can be shown that
\beq
\frac{\dot{B}}{B}=\frac{U^{\prime}}{R^{\prime}}=\frac{2\pi\rho r^{\frac{3}{2}}}{\sqrt{M/2}}-\frac{\sqrt{M/2}}{r^{\frac{3}{2}}}.
\eeq
Therefore, by definition, the initial profile of $K$ is given by
\beq
K(r)=\frac{2\pi\rho r^{\frac{3}{2}}}{\sqrt{M/2}}+\frac{3\sqrt{M/2}}{r^{\frac{3}{2}}}.
\eeq
Following this procedure we can obtain all the initial conditions needed for our system of 
equations.

The boundary conditions (at $r\to 0$ and $r\to \infty$) are easy to impose, and will not be listed here.

The formation of a black hole can be indicated by the appearance of a black hole apparent horizon, on which we have
\beq
U+\Gamma = 0,~~~~~ U-\Gamma < 0
\eeq
in our coordinates. Then the black hole mass can be read as the Misner-Sharp mass on the horizon, $M=R/2$. More details can be found in Ref. \cite{Deng:2017uwc} and references therein.

\end{document}